# Laser Probing of Neutron-Rich Nuclei in Light Atoms

Z.-T. Lu

*Physics Division, Argonne National Laboratory, Lemont, Illinois 60439, USA and*
*Department of Physics and Enrico Fermi Institute, University of Chicago, Chicago, Illinois 60637, USA*

P. Mueller

*Physics Division, Argonne National Laboratory, Lemont, Illinois 60439, USA*

G. W. F. Drake

*Department of Physics, University of Windsor, Windsor, Ontario N9B 3P4 Canada*

W. Nörtershäuser

*Institut für Kernphysik, Technische Universität Darmstadt, 64289 Darmstadt, Germany*

Steven C. Pieper

*Physics Division, Argonne National Laboratory, Argonne, Illinois 60439, USA*

Z.-C. Yan

*State Key Laboratory of Magnetic Resonance and Atomic and Molecular Physics, Wuhan Institute of Physics and Mathematics, and Center for Cold Atom Physics, Chinese Academy of Sciences, Wuhan 430071, China and*
*Department of Physics, University of New Brunswick, Fredericton, New Brunswick E3B 5A3 Canada*

The neutron-rich $^6$He and $^8$He isotopes exhibit an exotic nuclear structure that consists of a tightly bound $^4$He-like core with additional neutrons orbiting at a relatively large distance, forming a halo. Recent experimental efforts have succeeded in laser trapping and cooling these short-lived, rare helium atoms, and have measured the atomic isotope shifts along the $^4$He - $^6$He - $^8$He chain by performing laser spectroscopy on individual trapped atoms. Meanwhile, the few-electron atomic structure theory, including relativistic and QED corrections, has reached a comparable degree of accuracy in the calculation of the isotope shifts. In parallel efforts, also by measuring atomic isotope shifts, the nuclear charge radii of lithium and beryllium isotopes have been studied. The techniques employed were resonance ionization spectroscopy on neutral, thermal lithium atoms and collinear laser spectroscopy on beryllium ions. Combining advances in both atomic theory and laser spectroscopy, the charge radii of these light halo nuclei have now been determined for the first time independent of nuclear structure models. The results are compared with the values predicted by a number of nuclear structure calculations, and are used to guide our understanding of the nuclear forces in the extremely neutron-rich environment.

**Contents**

## I. INTRODUCTION

The $^4$He nucleus, or $\alpha$-particle, is a stable and tightly bound nuclear system, to which an additional neutron cannot be attached. The would-be $^5$He nucleus is unbound; its resonance state has an energy width corre-sponding to a lifetime of $10^{-20}$ s (Firestone and Shirley, 1996). On the other hand, an $\alpha$-particle and two neutrons can form a $^6$He nucleus that is stable under the strong interaction and only unstable under the influence of the weak interaction; it decays by $\beta$ emission with a half-life of 0.8 s. Here, the pairing between the two additional neutrons and the three-body nuclear force plays an essential role in stabilizing the $^6$He nucleus, which can be viewed as a three-body system $\alpha$-n-n. If any one of the three constituents is removed, the remaining two bodies become unbound. This interesting property is analogous to the topological properties of the Borromean rings (Fig. 1), thus earning $^6$He the nickname "Borromean nucleus" (Zhukov *et al.*, 1993). The pairing of neutrons continues along the isotope chain; $^7$He is unbound, and $^8$He is bound with a half-life of 0.1 s due to $\beta$ decay. While a typical nucleus has a neutron-to-proton ratio in the range of 1–1.5, for $^8$He the ratio is 3. Indeed, $^8$He holds the highest neutron-to-proton ratio among all known nuclides. At the same time, $^6$He and $^8$He have some of the lowest two-neutron separation energies. Both $^6$He and $^8$He consist of a tightly bound $\alpha$-core with additional neutrons orbiting at a relatively large distance, forming a halo (Fig. 2).

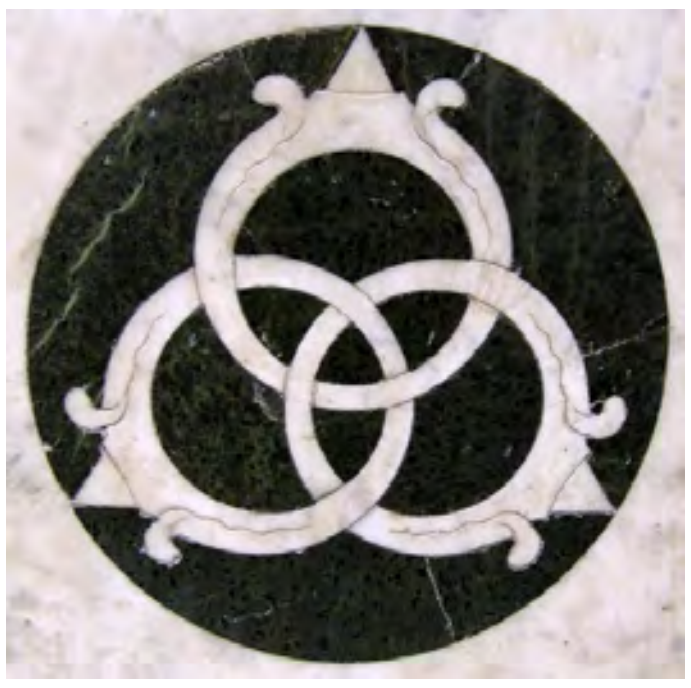

FIG. 1 Borromean rings depicted here as a marble inlay in the Church of San Pancrazio, Florence. The topology is such that the three rings are linked even though no two-ring pairs are linked. Analogies are found in certain nuclear bindings such as the $\alpha$-n-n structure of $^6$He.

These exotic nuclear phenomena are interesting to explore in their own right. Recent review articles have covered studies of halo structure in nuclei (Frederico *et al.*, 2012; Tanihata *et al.*, 2013) and in quantum systems in general (Jensen *et al.*, 2004). Moreover, they offer opportunities to study nuclear forces under extreme conditions in relatively simple systems (mass number $A < 10$), which in turn helps the development of effective models of nuclear forces that can be used to accurately describe nuclear structure, interactions and reactions. This paper reviews recent advances toward this goal in areas including nuclear theory, atomic theory, and laser trapping and probing of short-lived isotopes.

## II. NUCLEAR RADII

### A. Radii defined

The size of a nucleus is a fundamental property and, along with its binding and excitation energy, is used to probe the depth and range of the nuclear potential. Since the spatial distribution of the protons and neutrons may differ, a phenomenon that is particularly pronounced in halo nuclei, there are several ways to describe the nuclear size. For example, the rms charge radius ($r_c$) is defined as

$$r_c{}^2 = \frac{1}{Z}\int \rho_c(r) r^2 d^3r \tag{1}$$

where $\rho_c(r)$ is the nuclear charge density normalized to the number of protons, $Z$. This is the radius that is directly probed in atomic transition frequency measurements (see Section III). Alternatively, when only concerned with the spatial distribution of the protons, one can define the rms point-proton radius ($r_p$) as

$$r_p{}^2 = \frac{1}{Z}\int \rho_p(r) r^2 d^3r \tag{2}$$

where $\rho_p(r)$ is the density of the protons under the assumption that each proton is a point particle. In other words, only the center-of-mass of each proton is considered. This is a theoretical concept introduced for the benefit of not having to compute the size of the proton itself — a quantity beyond the realm of nuclear structure theories. Similarly, one can define the rms point-neutron radius ($r_n$), and the rms point-nucleon radius ($r_m$, or matter radius) where all nucleons, both neutrons and protons, are included. These are related by

$$r_m{}^2 = \frac{Z}{A} r_p{}^2 + \frac{N}{A} r_n{}^2 \tag{3}$$

where $N$ is the number of neutrons. The charge radius is related to the point-proton radius as ($\hbar = c = 1$) (Friar *et al.*, 1997)

$$r_c{}^2 = r_p{}^2 + \left(R_p{}^2 + \frac{3}{4M_p{}^2}\right) + \frac{N}{Z} R_n{}^2 + r_{\rm so}{}^2 + r_{\rm mec}{}^2 \tag{4}$$

Here $R_p$ is the charge radius of the proton itself; its value accepted by CODATA 2010 and the Particle Data Group (PDG) is 0.8775(51) fm (Beringer *et al.*, 2012; Mohr *et al.*, 2012), leading to $R_p^2 = 0.770(9)$ fm$^2$. The Darwin-Foldy term, $3/(4M_p{}^2) = 0.033$ fm$^2$, accounts for the charge distribution by virtual particle-antiparticle pairs that both surround and are polarized by the "bare" proton. Even a hypothetical point proton would have a nonzero charge radius due to the "Zitterbewegung" effect, a rapid oscillating motion of the proton resulting from the interference between the positive and negative frequency parts of the proton's wavefunction (Baym, 1969). One could argue that the Darwin-Foldy term should have been absorbed into the ${R_p}^2$ term to represent the apparent mean square charge radius of the proton (Friar *et al.*, 1997). Alas, the electron scattering community chose by convention to have these two terms separated so that ${R_p}^2$ represents the contribution purely due to the proton's "internal structure". The mean square charge radius of the neutron, ${R_n}^2$, has a PDG value of $-0.1161(22)$ fm$^2$. This negative value reflects the fact that the neutron consists of a positively charged core surrounded by negative charges on the outside. The spin-orbit term, ${r_{\rm so}}^2$, is due to the spin-orbit coupling of the nucleons with nonzero orbital angular momenta. It can be viewed as contribution to the charge density due to Lorentz boosting of anomalous magnetic moments (Ong *et al.*, 2010). Finally, ${r_{\rm mec}}^2$ represents the contribution of meson-exchange currents binding the nucleons and adding a small contribution to the nuclear charge density (Friar *et al.*, 1997).

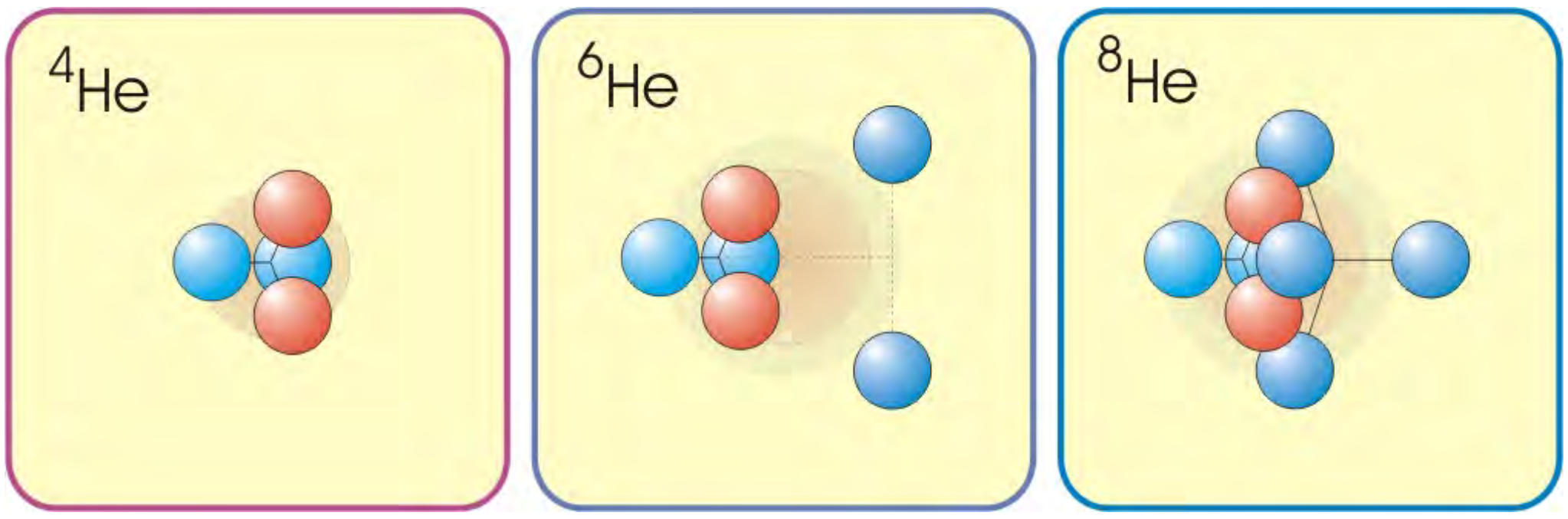

FIG. 2 Illustration of the nuclear structure of $^4$He, $^6$He and $^8$He. Red spheres represent protons; blue spheres neutrons. Red shadow indicates the area of motion of the protons; blue shadow of the neutrons. The nuclear charge radius is predominantly a measure of the center-of-mass motion of the charge carrying $^4$He-like core in $^6$He and $^8$He and depends on the correlation of the halo neutrons.

### B. Charge radii of $^4$He and of the proton

The charge radius of a stable, abundant isotope can be accurately determined using elastic scattering between a medium energy ($\sim 10^2$ MeV) electron beam and a target made of this isotope (Sick, 2001). The differential scattering cross section can be expressed as the formula of Mott scattering (for a point nucleus) modified by the form factor of the nucleus (Ottermann *et al.*, 1985).

$$\left(\frac{{\rm d}\sigma}{{\rm d}\Omega}\right)_{\rm exp} = \left(\frac{{\rm d}\sigma}{{\rm d}\Omega}\right)_{\rm Mott} F^2(q^2) \qquad (5)$$

where $q^2$ is the square of the four-momentum transfer during the collision. At $q^2$ of 0.04 GeV$^2$, the electron beam probes structure at a distance scale of $\sim$1 fm. In the case of $^4$He, with zero nuclear spin and no magnetic moment, the form factor is simply the Fourier transform of the charge density and can be expressed in a Taylor expansion as

$$F(q^2) = 1 - \frac{1}{3!}q^2 {r_c}^2 + \cdots \qquad (6)$$

For $^3$He or the proton, with nonzero magnetic moment, the form factor consists of both an electric and a magnetic part, which can be separated according to their dependencies on the scattering angle. The charge radius can then be extracted from the electric form factor as in Eq. (6) (Ottermann *et al.*, 1985).

The charge radii of $^4$He and $^3$He have been measured many times using the electron scattering method. For example, using an electron beam from the Mainz 400 MeV electron linear accelerator, and targets made of high-pressure gas cells, Ottermann *et al.* (1985) determined the charge radii to be 1.671(14) fm for $^4$He and 1.976(15) fm for $^3$He. A recent comprehensive study of the world data on elastic electron-helium scattering resulted in a more precise value of the charge radius of $^4$He, 1.681(4) fm (Sick, 2008).

The charge radius of the proton has also been measured in various electron scattering experiments. Recent measurements were performed at Jefferson Lab (Zhan *et al.*, 2011) using a 1.2 GeV polarized electron beam incident on a liquid hydrogen target and at the Mainz University Microtron MAMI (Bernauer *et al.*, 2010) with electron beam energies ranging from 180 to 855 MeV. The charge radius of the proton was determined to be 0.875(10) fm and 0.879(8) fm, respectively, consistent with the CODATA value 0.8775(51) fm (Mohr *et al.*, 2012), which is mainly inferred from atomic hydrogen spectroscopy.

Spectroscopy of muonic rather than electronic atoms is much more sensitive to nuclear charge radii because the muon is heavier than the electron and, thus, closer to the nucleus inside the atom. Recently, the Lamb shift of the 2s–2p transition in muonic hydrogen was measured for the first time at the cyclotron facility of Paul Scherrer Institute (PSI) (Pohl *et al.*, 2010). The new result for the charge radius of the proton, $R_p = 0.84087(39)$ fm, deviates from the CODATA value by $7\sigma$ (Antognini *et al.*, 2013). This, of course, has caused great excitement in the field, and motivated further work both in experiment and theory (Jentschura, 2011a,b). Muonic $^4$He ion has also been studied, but no transition sensitive to Lamb shift has been observed yet (Hauser *et al.*, 1992). A more sensitive experiment is being prepared at PSI (Antognini *et al.*, 2011).

Because of their needs for a macroscopic target, experiments of both electron scattering and muonic atom spectroscopy have so far been attempted on stable isotopes only, with the notable exception of tritium (Collard *et al.*, 1963). New techniques are being proposed and developed to perform electron scattering measurements on short-lived isotopes. The collaboration for electron-ion scattering experiments (ELISe) proposed to build an $eA$ collider between a circulating beam of electrons and exotic isotopes in a storage ring at the FAIR facility in Damstadt, Germany (Antonov *et al.*, 2011). In another scheme, named self-confining radioactive isotope target (SCRIT), Suda and Wakasugi (Suda and Wakasugi, 2005) proposed to have radioactive ions trapped directly by the colliding electron beam. In principle, an $eA$ collision could be arranged in inverse kinematics with a short-lived isotope beam scattering off a target containing electrons. However, the difficulty with this experimental design is that the momentum transfers are too low for measurements of the nuclear form factors and radii. At a beam energy of 0.7 GeV/u, the $q^2$ is only $6\times10^{-7}$ GeV$^2$ for $eA$ collisions.

For these reasons, the determination of nuclear charge radii for short-lived isotopes such as halo nuclei has not yet been possible, except by the isotope shift method discussed in section III. Therefore, it provides a unique measurement tool for this purpose.

## III. THEORY OF THE HELIUM ATOM

The measurement of nuclear sizes by the isotope shift method depends as much on accurate and reliable atomic structure calculations as it does on the isotope shift measurements themselves. This section discusses the relevant atomic states in question, and the theoretical methods used to calculate the mass-dependent contributions to the isotope shift. Figure 3 presents the helium atomic energy levels of interest. Laser excitation of helium atoms from the ground state requires vacuum ultraviolet photons at a wavelength of 58 nm — a region where precision lasers are not yet readily available, although much progress has been made recently in this area by using high-order harmonic generation of a frequency-comb laser (Cingoz *et al.*, 2012; Kandula *et al.*, 2011). Instead, most helium spectroscopy so far has been performed on the long-lived metastable states (Vassen *et al.*, 2012). In a neutral helium atom, the nucleus occupies a fractional volume on the order of $10^{-13}$, yet the minute perturbation on the atomic energy level due to the finite size of the nucleus can be precisely measured and calculated. Fig. 4(a) shows the electrostatic potential of a hypothetical point nucleus with zero charge radius. The electrostatic potential goes toward negative infinity as the electron approaches the nucleus at the origin. On the other hand, inside a real nucleus as depicted in Fig. 4(b), charge is distributed over the volume of the nucleus, and the electrostatic potential approaches a finite value at the origin. This effectively lifts the energy levels of the atomic states, with particularly significant results on the $s$–states whose electron wavefunctions do not vanish within the nucleus. For example, the transition frequencies of $2\,^3\mathrm{S}_1 - 3\,^3\mathrm{P}_\mathrm{J}$ in a helium atom are shifted down by a few MHz, or a fractional change of $10^{-8}$, due to the finite nuclear charge radius.

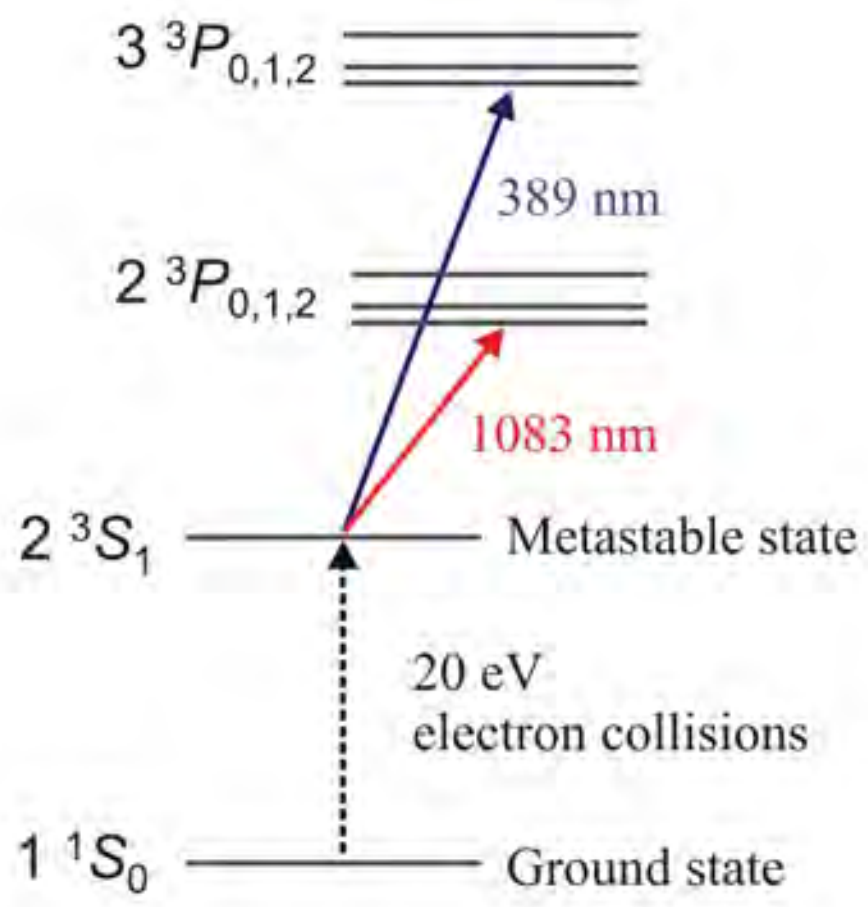

FIG. 3 The energy level diagram of the neutral helium atom. The $2\,^3\mathrm{S}_1$ state is metastable. Laser excitation on the $2\,^3\mathrm{S}_1$ - $2\,^3\mathrm{P}_2$ transition at 1083 nm was used to trap and cool helium atoms. Laser excitation on the $2\,^3\mathrm{S}_1$ - $3\,^3\mathrm{P}_J$ transition at 389 nm was used to detect the trapped atoms and measure their isotope shifts. Details are provided in Section IV.A.

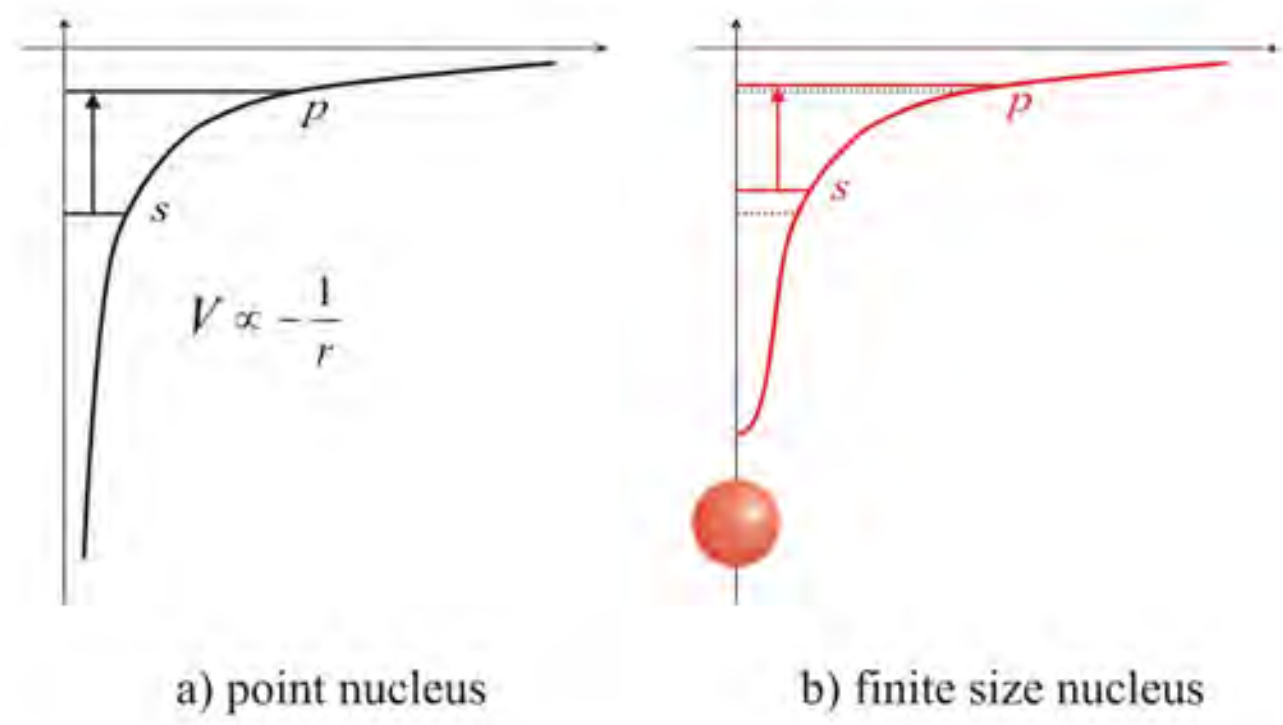

FIG. 4 The electrostatic potential and energy of bound $s$- and $p$-electronic levels are illustrated in a) for a hypothetical point nucleus, and in b) for the real case of a nucleus with a finite volume. The higher potential within the finite-sized nucleus causes the electrons to be less bound. This so-called *volume effect* is most pronounced for $s$-electrons.

This section covers the necessary high precision theory of the helium atom. In calculations and discussions, it is convenient to arrange the various contributions to the energy of the atom in the form of a double perturbation expansion in powers of the fine-structure constant $\alpha \simeq 1/137$ and the ratio of the reduced electron mass over the mass of the nucleus $\mu/M \simeq 10^{-4}$. Table I summarizes the various contributions to the energy, including the QED corrections and the finite nuclear size term. Since all the lower-order terms can now be calculated to very high precision, including the QED terms of order $\alpha^3$, the dominant source of uncertainty comes from the QED corrections of order $\alpha^4$ or higher. Yet, this QED uncertainty ($\sim$ 10 MHz) is larger than the finite nuclear size effect, thus preventing an extraction of the nuclear size directly from a single atomic transition frequency measurement. On the other hand, for the isotope shift, the QED terms independent of $\mu/M$ cancel, and so it is only the radiative recoil terms of order $\alpha^4\mu/M$ ($\sim$ 10 kHz) that contribute to the uncertainty. Since this is much less than the finite nuclear size correction of about 1 MHz, the comparison between theory and experiment clearly provides a means to determine the difference of the mean square radii between two isotopes of the same element. For example, the isotope shift of $^6$He-$^4$He for the transition $2\,^3{\rm S}_1 - 3\,^3{\rm P}_2$ is calculated to be (see Table III)

$$\mathrm{IS}_{6-4}(\mathrm{MHz}) = 43196.171(2) - 1.010[r_{c,6}^2 - r_{c,4}^2] \qquad (7)$$

TABLE I Contributions to the electronic binding energy and their orders of magnitude in atomic units. $a_0$ is the Bohr radius, $\alpha \approx 1/137$. For helium, the atomic number $Z = 2$, the mass ratio $\mu/M \sim 1 \times 10^{-4}$. $g_I$ is the nuclear g-factor. $\alpha_{\rm d}$ is the nuclear dipole polarizability.

| Contribution | Magnitude |
|---|---|
| Nonrelativistic energy | $Z^2$ |
| Mass polarization | $Z^2\mu/M$ |
| Second-order mass polarization | $Z^2(\mu/M)^2$ |
| Relativistic corrections | $Z^4\alpha^2$ |
| Relativistic recoil | $Z^4\alpha^2\mu/M$ |
| Anomalous magnetic moment | $Z^4\alpha^3$ |
| Hyperfine structure | $Z^3 g_I \mu_0^2$ |
| Lamb shift | $Z^4\alpha^3 \ln\alpha + \cdots$ |
| Radiative recoil | $Z^4\alpha^3(\ln\alpha)\mu/M$ |
| Finite nuclear size | $Z^4\langle r_c/a_0\rangle^2$ |
| Nuclear polarization | $Z^3 e^2 \alpha_{\rm d}/(\alpha a_0^4)$ |

The first term on the right-hand side, which dominates in the case of helium or other light nuclides, is referred to as the *mass shift* (due to electronic structure), and the second term as the *volume shift* (due to the finite nuclear volume). The radii are in the units of fm. In order to extract the nuclear charge radius of $^6$He from an isotope shift measurement, it is essential to have precise calculations of both the mass shift and the coefficient of the volume shift. This in turn requires the precisely determined masses of both isotopes.

Direct mass measurements of both $^6$He and $^8$He have been performed recently with the TITAN Penning trap mass spectrometer at the TRIUMF ISAC facility (Brodeur *et al.*, 2012; Ryjkov *et al.*, 2008). Once produced in a spallation reaction, $^6$He or $^8$He atoms were ionized, mass selected, and transported to the TITAN facility, where they were first thermalized and accumulated using a hydrogen-filled radio-frequency quadrupole (RFQ) ion trap, and then injected into a Penning trap for mass measurements. For $^6$He, the new result, 6.018 885 883(57) u (atomic mass unit), deviates from the previous value (AME03) by $4\sigma$, while improving the precision by a factor of 14. For $^8$He, the new result, 8.033 934 44(11) u, agrees with the previous value (AME03) within $1.7\sigma$, while improving the precision by a factor of 12. Prior to these direct measurements, the mass uncertainties would cause uncertainties in the radii at the 1% level. The new measurements are so precise that mass uncertainties have become negligible in the current extraction of nuclear charge radii.

### A. Solution to the nonrelativistic Schrödinger equation

The starting point for the calculation is to find accurate solutions to the nonrelativistic Schrodinger equation. This is the foundation upon which are built the relativistic and QED corrections by perturbation theory. The usual methods of theoretical atomic physics, such as the Hartree-Fock approximation or configuration interaction methods, are not capable of yielding results of spectroscopic accuracy. Hence, specialized techniques have been developed (Drake, 1993a,b; Drake and Yan, 1992).

Considering first the case of infinite nuclear mass, the nonrelativistic Hamiltonian for a two-electron atom is given by

$$H_\infty = -\frac{1}{2}\nabla_1^2 - \frac{1}{2}\nabla_2^2 - \frac{Z}{r_1} - \frac{Z}{r_2} + \frac{1}{r_{12}} \qquad (8)$$

where atomic units are used with $\hbar = m_e = e = 1$ and $-i\nabla_{1,2}$ representing the momentum operators. Because of the electron-electron repulsion term $1/r_{12}$, with $r_{12} = |\mathbf{r}_1 - \mathbf{r}_2|$ being the inter-electronic separation, the Hamiltonian is nonseparable, and so the Schrödinger equation cannot be solved exactly. As long ago as 1929, Hylleraas (Hylleraas, 1929) suggested expanding the wave function in an explicitly correlated variational basis set of the form (in modern notation)

$$\Psi_\infty(\mathbf{r}_1, \mathbf{r}_2) = \sum_{i,j,k} a_{ijk}\, r_1^i r_2^j r_{12}^k\, e^{-\alpha r_1 - \beta r_2}\, \mathcal{Y}_{l_1 l_2 L}^M(\hat{\mathbf{r}}_1, \hat{\mathbf{r}}_2) \qquad (9)$$

where $\mathcal{Y}_{l_1 l_2 L}^M$ is a vector-coupled product of spherical harmonics to form a state of total angular momentum $L$ and component $M$. The coefficients $a_{ijk}$ are linear variational parameters; $\alpha$ and $\beta$ are nonlinear variational parameters that set the distance scale for the wave function. As shown by Klahn and Bingel (Klahn and Bingel, 1977, 1978), the basis set is complete in the limit that the number of powers tends to infinity. This important property ensures that the results converge to the correct answer, including all correlation effects. In the calculation, $\alpha$ and $\beta$ are separately optimized for each set of angular momentum terms. For excited states, it is desirable to double the basis set so that each combination of powers $\{i, j, k\}$ is included two (or more) times with different values $\alpha_1$, $\beta_1$, $\alpha_2, \beta_2$ etc. in different blocks (Drake *et al.*, 2002). This optimization produces a natural partition of the basis set into distinct sectors representing the asymptotic and short-range parts of the wave function.

The most studied example is the $1s^2\,{}^1\mathrm{S}_0$ ground state of helium. Convergence to 20 or more figures can be readily obtained using conventional quadruple precision (32 decimal digit) arithmetic (Drake *et al.*, 2002). Recently, even higher accuracy has been obtained by Schwartz (Schwartz, 2006), and by Nakashima and Nakatsuji (Nakashima and Nakatsuji, 2008). High precision results for all states of helium up to $n = 10$ and angular momentum $L = 7$ are available in Refs. (Drake, 1993a; Drake and Yan, 1992, 1994). Combined with asymptotic expansion methods for high $L$ and quantum defect methods for high $n$ (Drake, 1994), this provides a complete coverage of the entire spectrum of singly-excited states for helium.

### B. Mass, relativistic and QED corrections

In a calculation of the isotope shift, it is necessary to include also the motion of the nucleus. A transformation to the center-of-mass frame plus relative coordinates results in the mass polarization term $-(\mu/M)\nabla_1 \cdot \nabla_2$ to be added to $H_\infty$ (Bethe and Salpeter, 1957). Its effect is calculated up to the second order $(\mu/M)^2$.

Relativistic corrections corresponding to the term of order $\alpha^2 Z^4$ (Table I) come from the nonrelativistic form of the Breit interaction (Bethe and Salpeter, 1957). There are also relativistic recoil terms of the order $\alpha^2 Z^4 \mu/M$ (Drake, 1993a), arising from the mass scaling of the terms in the Breit interaction, a transformation to center-of-mass coordinates (Stone, 1961, 1963), and mass-dependent corrections to the wave function due to the mass polarization term in the Hamiltonian.

Accurate calculations of QED corrections to the lowest order $\alpha^3$ and $\alpha^3\mu/M$ once presented a major limitation on the accuracy that could be achieved in atomic structure calculations. However, this problem has now been solved, (see for example Ref. (Drake and Goldman, 2000)), and higher order corrections can be estimated from combinations of known hydrogenic results (Yan and Drake, 1998). The orders of magnitude for higher-order corrections are discussed by Drake and Martin (Drake and Martin, 1999) for helium. Their contribution to the transition energy is taken to be the QED uncertainty. A comprehensive tabulation of energy levels for $^3$He and $^4$He, including hyperfine structure, has been given by Morton *et al.* (Morton *et al.*, 2006). Further improved calculations on hyperfine structure by Pachucki *et al.* (Pachucki *et al.*, 2012) are in excellent agreement with experiment.

TABLE II Contributions to the isotope shifts in the ionization energies of $^6$He relative to $^4$He. Units are MHz. The term $\alpha^3\mu/M$, due to QED correction, includes estimates of the higher-order terms.

| Term | $2\,{}^3\mathrm{S}_1$ | $2\,{}^3\mathrm{P}_2$ | $3\,{}^3\mathrm{P}_2$ |
|---|---|---|---|
| $\mu/M$ | 55 195.486(2) | 20 730.132(1) | 12 000.665(1) |
| $(\mu/M)^2$ | –3.964 | –14.132 | –4.847 |
| $\alpha^2\mu/M$ | 1.435 | 3.285 | 0.724 |
| $\alpha^3\mu/M$ | –0.280 | –0.206 | –0.036 |
| $\Delta E_{\mathrm{pol}}$ | 0.0157(28) | –0.0048(9) | –0.0014(2) |
| Total | 55 192.693(3) | 20 719.074(1) | 11 996.505(1) |

### C. Atomic isotope shifts

The results of the previous section can now be assembled to calculate the total mass-dependent contribution to the total isotope shift for the transition in question. This is the quantity that must be subtracted from the measured isotope shift (see Eq. 7) in order to isolate the nuclear volume effect. As examples, the various contributions to the isotope shift are listed in Table II for $^6$He relative to $^4$He. The results are expressed as contributions to the isotope shift for the ionization energy of each state so that the isotope shift for the transition is obtained by subtracting the entries for the corresponding initial and final states. The terms are classified according to their dependence on $\mu/M$ and $\alpha$, as given in Table I. The term of order $(\mu/M)^2$ comes from second-order mass polarization. The relativistic recoil terms of order $\alpha^2\mu/M$ come from mass scaling, mass polarization, and the Stone terms (Stone, 1961, 1963). The radiative recoil terms similarly come from a combination of mass scaling, mass polarization, and higher-order recoil corrections, as discussed by Pachucki and co-workers (Pachucki and Sapirstein, 2003). In addition, Puchalski *et al.* (Puchalski *et al.*, 2006) have discussed a correction to the isotope shift due to nuclear polarizability. The correction is given by

$$\Delta E_{\mathrm{pol}} = -mc^2\alpha \sum_i \langle \delta(\mathbf{r}_i) \rangle \alpha_{\mathrm{d}} \qquad (10)$$

where $\alpha_{\mathrm{d}}$ is an averaged nuclear dipole polarizability (Puchalski *et al.*, 2006). For the case of $^6$He ($\alpha_{\mathrm{d}} = 24.7 \pm 5.0$ fm$^3$), this gives significant additional contributions to the isotope shifts (Pachucki and Moro, 2007). The correction is negligibly small for $^8$He. Finally, the correction for the finite nuclear size is given in the lowest

TABLE III Parameters for the determination of nuclear radii from the measured isotope shift (see Eq. (12)). The uncertainties in the mass shifts are due to the uncertainties in atomic masses, and higher-order finite mass corrections are not included in the calculation. The mass shift values do not include the nuclear polarization correction, which is –0.014(3) MHz for $^6$He and –0.002(1) MHz for $^8$He.

| Isotopes | Transition | $\delta\nu_{\rm MS}$ (MHz) | $K_{\rm i,f}$ (MHz/fm$^2$) |
|---|---|---|---|
| $^6$He – $^4$He | $2\,^3$S$_1 - 3\,^3$P$_0$ | 43196.1573(16) | 1.0104(1) |
| | $2\,^3$S$_1 - 3\,^3$P$_1$ | 43195.8966(16) | 1.0104(1) |
| | $2\,^3$S$_1 - 3\,^3$P$_2$ | 43196.1706(16) | 1.0104(1) |
| $^8$He – $^4$He | $2\,^3$S$_1 - 3\,^3$P$_0$ | 64702.4888(18) | 1.0108(1) |
| | $2\,^3$S$_1 - 3\,^3$P$_1$ | 64702.0982(18) | 1.0108(1) |
| | $2\,^3$S$_1 - 3\,^3$P$_2$ | 64702.5086(18) | 1.0108(1) |

order by

$$\Delta E_{\rm nuc} = \frac{2\pi Z e^2 r_{\rm c}^2}{3}\sum_i \langle \delta^3(\mathbf{r}_i)\rangle \tag{11}$$

Because of the dependence on $r_{\rm c}^2$, the measured isotope shift $\delta\nu$ for a transition i $\rightarrow$ f is then related to the calculated mass shift $\delta\nu_{\rm MS}$ by an equation of the form

$$\delta\nu = \delta\nu_{\rm MS} - K_{\rm i,f}\left[r_{{\rm c},A}^2 - r_{{\rm c},B}^2\right] \tag{12}$$

and $K_{\rm i,f} = (2\pi Z e^2/3)\sum_j[\langle\delta^3(\mathbf{r}_j)\rangle_{\rm i} - \langle\delta^3(\mathbf{r}_j)\rangle_{\rm f}]$ is nearly independent of the particular isotopes $A$ and $B$. The calculated values of the parameters $\delta\nu_{\rm MS}$ and $K_{\rm i,f}$ are listed in Table III for the transitions of interest.

Table III indicates that the coefficient for the volume shift is constant at the $10^{-3}$ level among the fine-structure triplet of transitions between $2\,^3$S$_1$ and $3\,^3$P$_{\rm J}$ ($\rm J = 0, 1, 2$), and for both the $^6$He–$^4$He and $^8$He–$^4$He isotope shifts. On the other hand, the mass shifts differ greatly between $^6$He–$^4$He and $^8$He–$^4$He. Interestingly, the mass shift for the $2\,^3$S$_1 - 3\,^3$P$_1$ transition stands out among the fine-structure triplet. This is caused by state mixing between $3\,^3$P$_1$ and $3\,^1$P$_1$ due to spin-orbit coupling. This peculiar effect has been verified in the case of $^6$He–$^4$He where the isotope shifts were measured for all three transitions between $2\,^3$S$_1$ and $3\,^3$P$_{\rm J}$ (Mueller *et al.*, 2007) (see Section IV).

## IV. LASER TRAPPING AND PROBING

The main challenge for laser spectroscopy of $^6$He and $^8$He is the combination of requirements on sensitivity and precision. In addition to the short half-lives and low yields of these two isotopes, the efficiency of placing the atoms into the metastable state $2\,^3$S$_1$ using electron impact excitation by a discharge is only $10^{-5}$. On the precision side, a 100 kHz uncertainty in the isotope shift results in approximately a 1% uncertainty in the charge radius. In order to meet all these challenges, laser trapping and cooling of $^6$He and $^8$He atoms was employed, and laser spectroscopy on individual atoms in the trap realized (Mueller *et al.*, 2007; Wang *et al.*, 2004). The selective cooling and trapping of helium atoms in a magneto-optical trap (MOT) was pivotal for this work, providing single-atom sensitivity, large signal-to-noise ratios and high spectroscopic resolution. In addition, the selectivity of the MOT guarantees that the trapped sample is absolutely free of any contamination by the dominant $^4$He isotope or any other atomic and molecular background (Chen *et al.*, 1999). The technique was developed at Argonne National Laboratory, and was first applied to laser spectroscopy of $^6$He at Argonne's ATLAS accelerator facility (Wang *et al.*, 2004). Later, following further developments, the apparatus was moved to the GANIL accelerator facility where an improved measurement on $^6$He and the first measurement on $^8$He were carried out (Mueller *et al.*, 2007).

### A. Trapping and probing of $^{6,8}$He

At GANIL, $^6$He and $^8$He were simultaneously produced via spallation from a primary beam of 75 MeV/u $^{13}$C impinging on a heated (2000 K) graphite target. Mass selected, low-energy (20 keV) beams of either $^6$He or $^8$He with yields of around $1\times10^8$ and $5\times10^5$ ions per second, respectively, were delivered to an adjacent low-radiation area (Landré-Pellemoine *et al.*, 2002) where the helium ion beam was stopped in a hot graphite foil for neutralization and fast release. Neutral, thermal helium atoms were subsequently compressed by a turbopump within 0.25 s into the atomic beam apparatus at the rates of approximately $5\times10^7$ s$^{-1}$ and $1\times10^5$ s$^{-1}$ for $^6$He and $^8$He, respectively.

Figure 5 provides the schematic of the atomic beam and trap apparatus. A beam of metastable helium atoms with a probable velocity around 1000 m/s was produced through a liquid-nitrogen-cooled gas discharge. Transverse cooling and Zeeman slowing were applied to capture the metastable helium atoms of a selected isotope into the magneto-optical trap (MOT). Cooling and trapping were based on repeated excitation of the cycling transition $2\,^3$S$_1 - 2\,^3$P$_2$ at the wavelength of 1083 nm (Fig. 3). Detection and spectroscopy of the atoms captured in the MOT were performed by exciting one of the three transitions, $2\,^3$S$_1 - 3\,^3$P$_{\rm J}$, at 389 nm and imaging the fluorescence light onto a photomultiplier tube. The signal-to-noise ratio of a single trapped atom reached 10 within 50 ms of integration time for photon counts. The total capture efficiency was $1\times10^{-7}$. When trapping $^6$He, there were typically a few $^6$He atoms in the trap, yielding a capture rate of around 20 000 $^6$He atoms per hour. On the other hand, when trapping $^8$He, single $^8$He atoms were captured at the rate of 30 per hour with each staying in the trap for an average time of 0.1 s. Samples of resonance peaks for $^8$He are given in Fig. 6, including

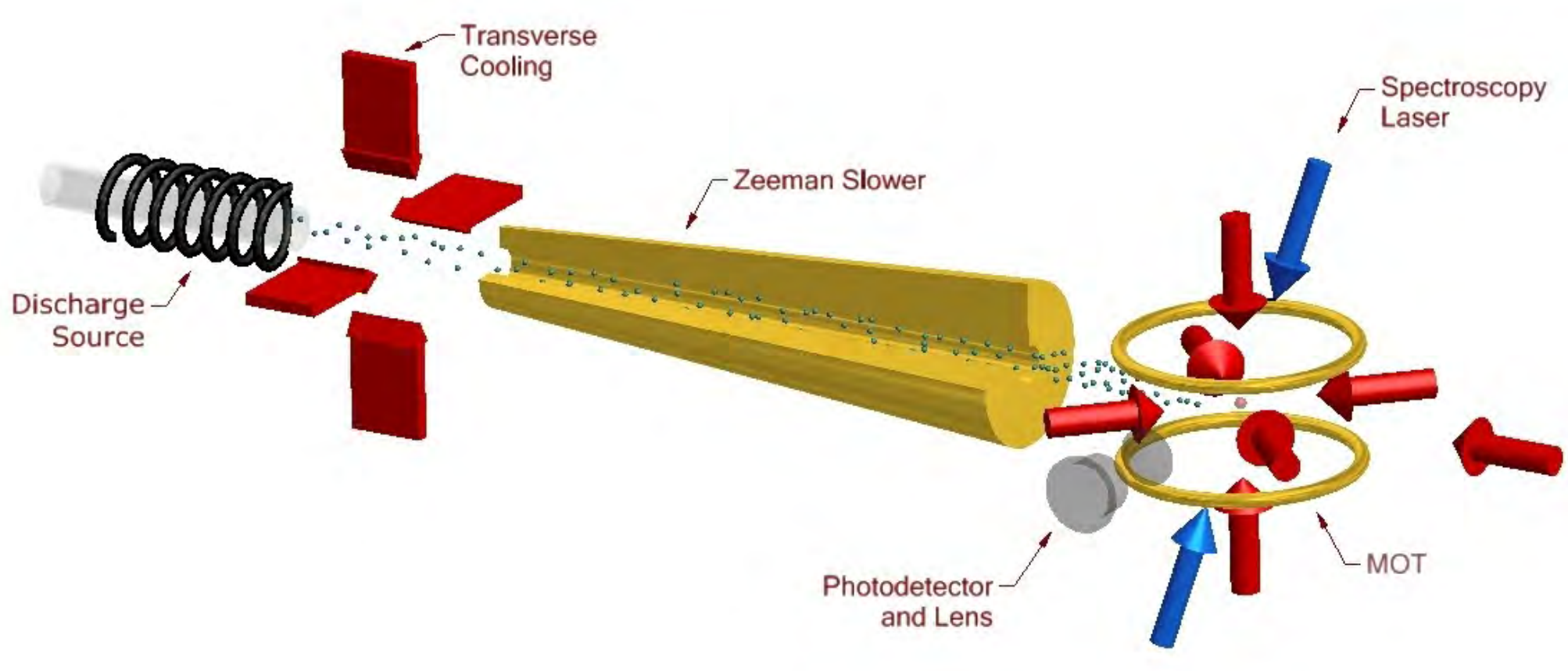

FIG. 5 Schematic of the $^{6,8}$He trap apparatus. A beam of metastable helium atoms is provided by a gas discharge source. Subsequently, the atoms are collimated through transverse cooling, decelerated in a Zeeman slower and captured by a magneto-optical trap (MOT). Light from the spectroscopy laser beams that is scattered by the trapped atoms is imaged onto a photon detector. More details are provided in Ref. (Wang *et al.*, 2004)

the spectrum of the very first $^{8}$He atom observed in the trap. It stayed in the trap for a notably long duration of 0.4 s.

The isotope shifts for $^{6}$He and $^{8}$He relative to $^{4}$He, obtained in the individual measurements, are plotted in Fig. 7 along with the extracted field shifts. Table IV lists the weighted averages of the isotope and field shifts separately for the different fine structure levels $^{3}$P$_{\rm J}$. The isotope shift values for the different transitions in $^{6}$He show variations by 250 kHz, as predicted by the atomic theory calculations. The extracted field shifts for all three transitions agree well within statistical uncertainties. This is a valuable consistency test for atomic theory as well as a check for a class of systematic errors in the experiment, since the strengths of these three transitions vary by up to a factor of five. Hence, the field shifts over all three transitions in $^{6}$He were averaged as independent measurements, and likewise for the two transitions observed in $^{8}$He.

The final field shift results for both isotopes are listed in Table IV along with the contributions from statistical and systematic uncertainties. A significant systematic uncertainty is caused by Zeeman shifts that might have varied among isotopes if the atoms were not located exactly at the zero $B$-field position of the MOT. Limits on this effect are set conservatively at $\leq 30$ kHz for the $^{6}$He–$^{4}$He isotope shift, and $\leq 45$ kHz for $^{8}$He–$^{4}$He.

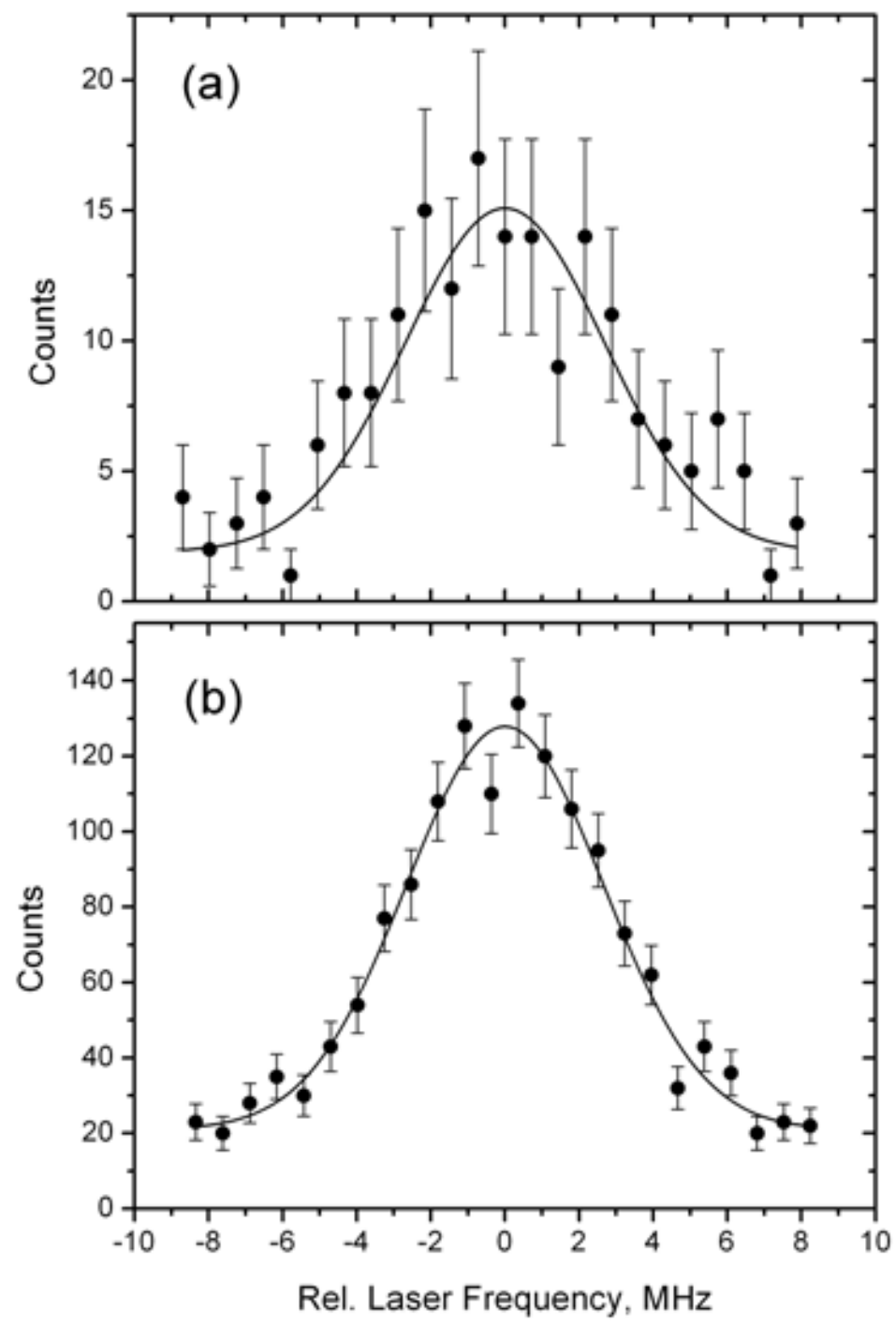

FIG. 6 Sample spectra for $^{8}$He taken on the $2\,^{3}$S$_1 - 3\,^{3}$P$_2$ transition at a probing laser intensity of $\sim 3 \times I_{\rm sat}$. Error bars are statistical uncertainties, and the lines represent least squares fits using Gaussian profiles. (a) Spectrum of a single $^{8}$He atom, the very first observed in the trap. It stayed in the trap for an extra long time of 0.4 s. The fit results in a statistical frequency uncertainty of 320 kHz with $\chi^2 = 0.84$. (b) Spectrum accumulated over 30 trapped atoms. Uncertainty is 110 kHz with $\chi^2 = 0.87$.

### B. Study of systematic uncertainties with $^{3,4}$He

Studies of $^{3,4}$He atoms were carried out with the same apparatus used for $^{6,8}$He as a check for systematic effects, but the results are interesting in their own right. A small atomic cloud consisting of tens of $^{4}$He atoms was loaded into the MOT and the $3\,^{3}$P$_{0,1,2}$ fine structure intervals were measured (Mueller *et al.*, 2005). The standard devi-ation of 30 measurements under different trap conditions was 40 kHz, which represents the systematic uncertainty due to trap effects. Meanwhile, the same fine structure was studied by performing laser spectroscopy on a collimated atomic beam. The experimental results from both the trap and atomic beam methods show excellent agreement with each other and with the theoretical calculation (Mueller *et al.*, 2005; Yan and Drake, 1994).

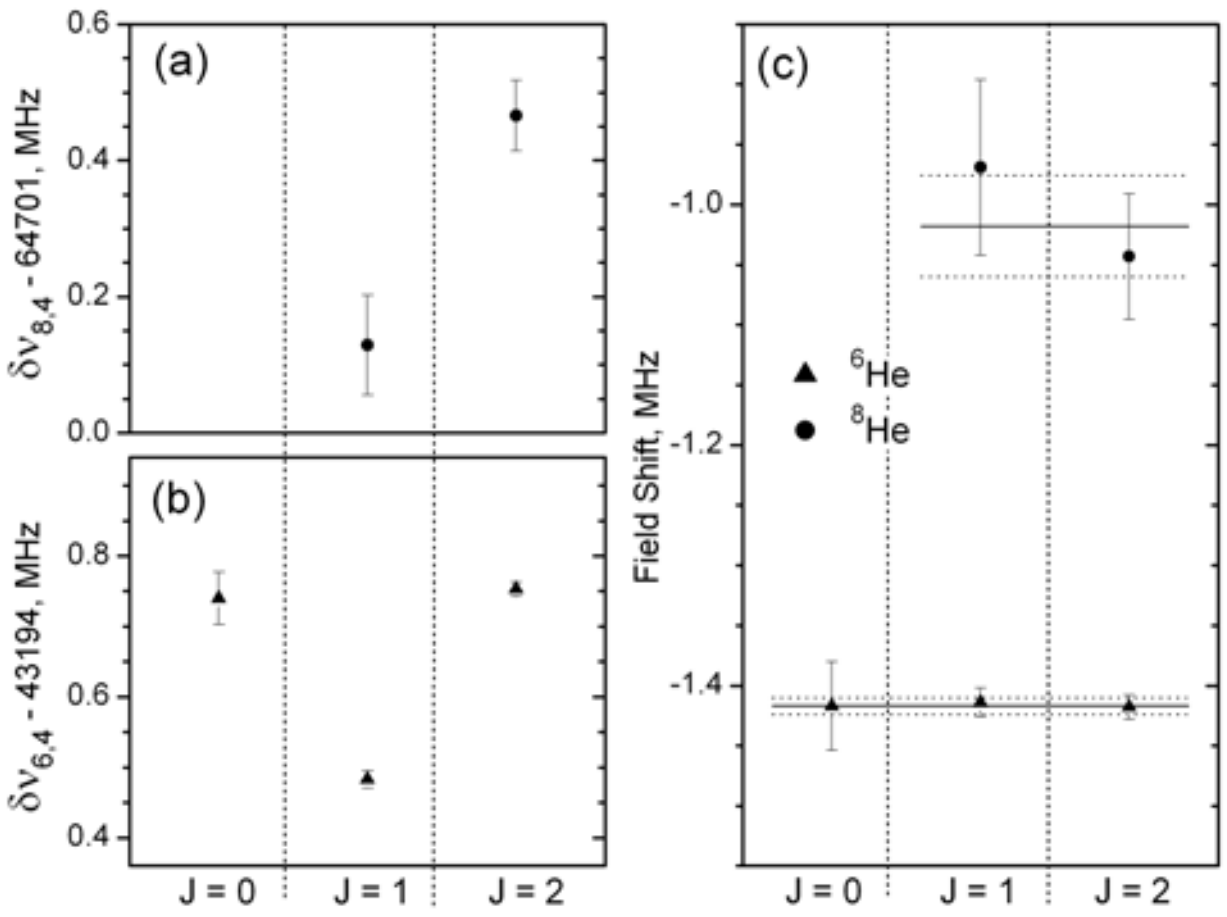

FIG. 7 Experimental isotope shifts relative to $^4$He from the individual measurements for $^8$He (a) and $^6$He (b). As expected, the total isotope shift depends on the J of the upper $3\,^3\mathrm{P_J}$ state. The extracted field shift values plotted in (c) show no systematic J dependence for either isotope. The horizontal lines in (c) mark the weighted averages and statistical error bands of the field shift.

TABLE IV Weighted averages of the experimental isotope shifts $\delta\nu$ (including recoil correction) for the different transitions in $^6$He and $^8$He from (a) (Mueller *et al.*, 2007) and (b) (Wang *et al.*, 2004). The field shift $\delta\nu^{\mathrm{FS}}_{A,4} = K_{\mathrm{i,f}}\ \delta\langle r^2\rangle_{A,4}$ was calculated for each transition using the mass shift values listed in Tab. III. The errors given in parentheses are the uncorrelated uncertainties. The listed weighted average of the field shift includes the nuclear polarization correction of –0.014(3) MHz for $^6$He and –0.002(1) MHz for $^8$He. The value given in square brackets denotes the common systematic uncertainty. All values are in MHz.

| | Transition | $\delta\nu_{A,4}$ (MHz) | $\delta\nu^{\mathrm{FS}}_{A,4}$ (MHz) | Ref. |
|---|---|---|---|---|
| $^6$He | $2\,^3\mathrm{S}_1 - 3\,^3\mathrm{P}_0$ | 43194.740(37) | 1.417(37) | (a) |
| | $2\,^3\mathrm{S}_1 - 3\,^3\mathrm{P}_1$ | 43194.483(12) | 1.414(12) | (a) |
| | $2\,^3\mathrm{S}_1 - 3\,^3\mathrm{P}_2$ | 43194.751(10) | 1.420(10) | (a) |
| | $2\,^3\mathrm{S}_1 - 3\,^3\mathrm{P}_2$ | 43194.772(33) | 1.399(33) | (b) |
| | weighted avg. + nucl. polarization | | 1.430(7)[30] | |
| $^8$He | $2\,^3\mathrm{S}_1 - 3\,^3\mathrm{P}_1$ | 64701.129(73) | 0.969(73) | (a) |
| | $2\,^3\mathrm{S}_1 - 3\,^3\mathrm{P}_2$ | 64701.466(52) | 1.043(52) | (a) |
| | weighted avg. + nucl. polarization | | 1.020(42)[45] | |

The measurement of the isotope shift between $^3$He and $^4$He in the trap would in principle provide the best test of the systematic effects in isotope shifts. However, the hyperfine structure of $^3$He complicates the situation. After subtracting the hyperfine shifts, the isotope shift between $^4$He and $^3$He in the $2\,^3\mathrm{S}_1 - 3\,^3\mathrm{P}_2$ transition is derived to be 42184.268(40)(100) MHz, where the first error is due to the measurement, and the second is from the uncertainty in the hyperfine shift calculation (Yan and Drake, 1994). This result agrees with the calculated value of 42184.299(5) using the input of the $^3$He charge radius (Shiner *et al.*, 1995). It also agrees with an earlier, less precise measurement (Marin *et al.*, 1995).

More recently, a measurement was made on the $^3$He–$^4$He isotope shift for the $2\,^3\mathrm{S}_1$–$2\,^3\mathrm{P_J}$ manifold of transitions (Cancio Pastor *et al.*, 2012). Together with an improved calculation to terms of order $m\alpha^6$ (Pachucki *et al.*, 2012), it provides the difference in rms charge radii $\delta r_c^2 = 1.066(4)\mathrm{fm}^2$. Given the charge radius of $^4$He (Sick, 2008), the charge radius of $^3$He is derived to be 1.973(4) fm. This new result is more accurate than and consistent with the earlier one, 1.976(15) fm, obtained using the electron scattering method (Ottermann *et al.*, 1985). There is also an isotope shift measurement in the highly forbidden $2\,^3\mathrm{S}_1$–$2\,^1\mathrm{S}_0$ relativistic M1 transition (van Rooij *et al.*, 2011). The result, following a reevaluation by Cancio Pastor *et al.* (Cancio Pastor *et al.*, 2012), is $\delta r_c^2 = 1.028(11)\ \mathrm{fm}^2$. The two recent measurements resulted in two values of $\delta r_c^2$ that differ by about $3\sigma$. The discrepancy is yet to be resolved.

Interestingly, investigations of the transitions in $^3$He led to the discovery of anomalous line strengths when compared with estimates from simple L–S coupling (Sulai *et al.*, 2008). The strengths of two "allowed transitions", $2\,^3\mathrm{S}_1$ F = 1/2 – $3\,^3\mathrm{P}_2$ F = 3/2 and $2\,^3\mathrm{S}_1$ F = 3/2 – $3\,^3\mathrm{P}_1$ F = 3/2, were found to be 1000 times weaker than that of the strongest transition in the same manifold, $2\,^3\mathrm{S}_1$ F = 3/2 – $3\,^3\mathrm{P}_2$ F = 5/2. This dramatic suppression of transition strengths is due to a rare atomic phenomenon; within the $3\,^3\mathrm{P_J}$ manifold, the hyperfine interaction is comparable to or even stronger than the fine-structure interaction. Consequently, the conventional model based on L–S coupling is no longer applicable. Rather, an alternative model, referred to as I–S coupling (Sulai *et al.*, 2008), where the fine-structure interaction is treated as a perturbation on states obtained by first coupling the nuclear spin to the total electron spin, provides a good qualitative explanation of the observed suppression.

## V. CHARGE RADII AND POINT-PROTON RADII OF $^{6,8}$HE

Combining the isotope shift measurements (Sec. IV.A), the independent determination of the charge radius of $^4$He (Sec. II.B), and the modern theory of the helium atom (Sec. III), the charge radii of $^6$He and $^8$He can be extracted in a way that is independent of nuclear models. The results are found in Table V.

TABLE V Experimental and theoretical charge and point-proton radii $r_c$ and $r_p$, binding energies $E_B$, and two-neutron separation energies $E_{2n}$ of helium isotopes. The experimental $r_c$ for $^6$He and $^8$He are extracted from the averaged isotope shifts of all observed transitions and the listed value of $r_c$ of $^4$He obtained from electron scattering. The experimental $r_p$ values are calculated following Eq. (4) using the PDG values for $R_n$ and $R_p$ (Beringer *et al.*, 2012) and $r_{\rm so}$ from (Papadimitriou *et al.*, 2011). The meson-exchange term $r_{\rm mec}$ is neglected. Units of radii are fm and of energies are MeV.

| Quantity | $^4$He | $^6$He | $^8$He | Ref. |
|---|---|---|---|---|
| $r_c$, e-scattering | 1.681(4) | | | a) |
| $r_c$, isotope shift | | 2.059(8) | 1.958(16) | b) |
| $r_p$, expt. | 1.462(6) | 1.934(9) | 1.881(17) | |
| $r_p$, AV18+IL7 | 1.432(3) | 1.92(3) | 1.83(2) | c) |
| $r_p$, JISP16 | 1.436(1) | 1.85(5) | 1.80(5) | d) |
| $E_B$, expt. | 28.30 | 29.27 | 31.41 | |
| $E_B$, AV18+IL7 | 28.43(1) | 29.20(3) | 31.06(15) | c) |
| $E_B$, JISP16 | 28.299(1) | 28.80(5) | 29.9(1) | d) |
| $E_{2n}$, expt. | – | 0.98 | 2.13 | e) |
| $E_{2n}$, AV18+IL7 | – | 0.97(3) | 1.86(15) | |
| $E_{2n}$, JISP16 | – | 0.50(5) | 1.1(1) | |

a) (Sick, 2008), b) This work, c) (Brida *et al.*, 2011) and this work, d) (Maris, 2013), e) (Wang *et al.*, 2012)

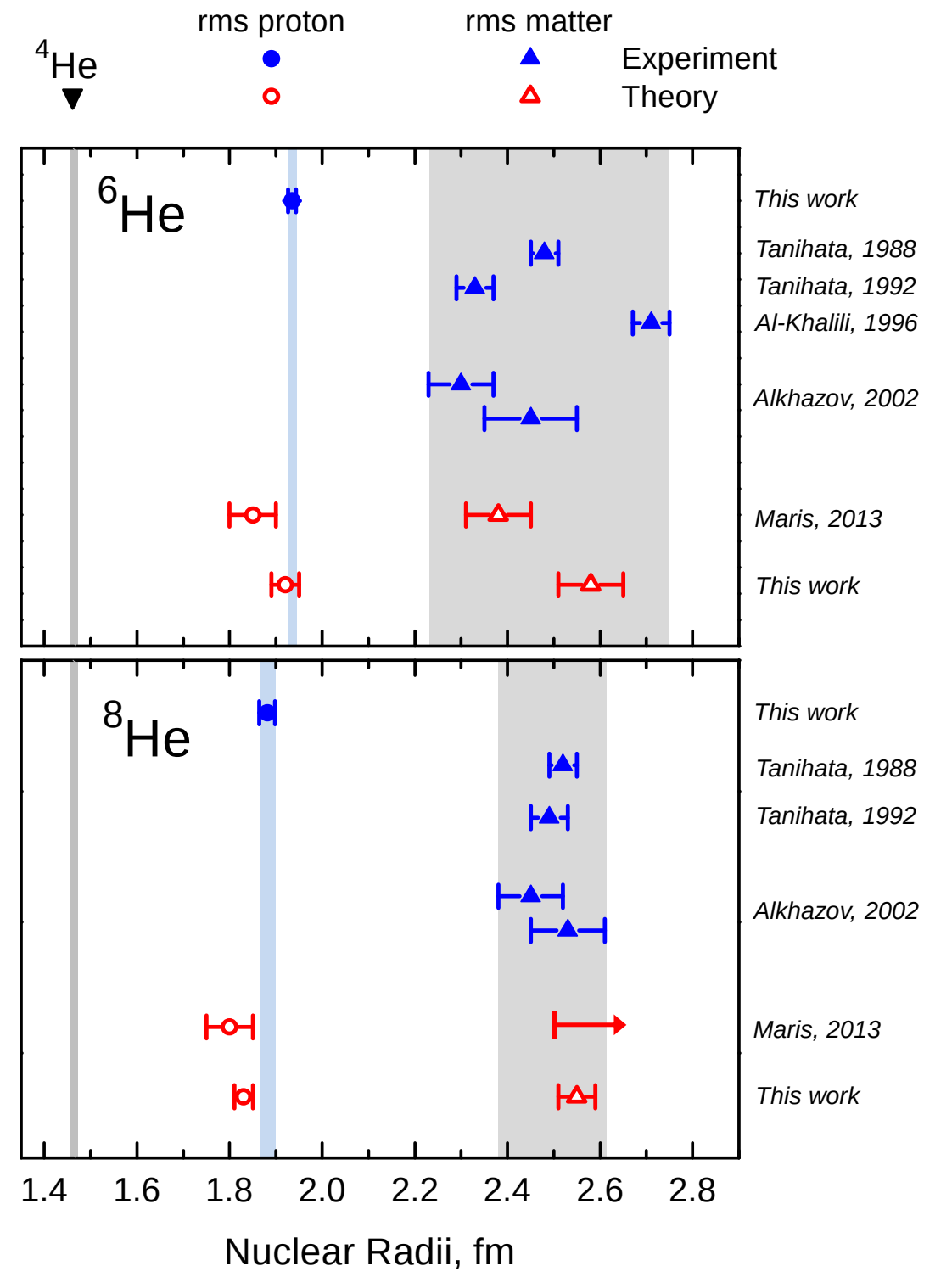

FIG. 8 Comparison between experimental and theoretical values of point-proton and matter radii for $^6$He (top panel) and $^8$He (bottom panel) (see also Tables VI and V).

In order to compare the results with values obtained from nuclear theory, a conversion needs to be performed between the charge radius and the point-proton radius following Eq. (4). Here, the 2012 PDG values of the neutron and proton charge radii (Beringer *et al.*, 2012) were used. The spin-orbit correction term was calculated in the framework of Gamow shell model (Papadimitriou *et al.*, 2011). This value is close to the spin-orbit term obtained in for the case of the halo neutrons in pure p-orbitals (Ong *et al.*, 2010).

The point-proton radii from this work are plotted in Fig. 8 along with matter radii extracted from scattering experiments (Table VI). While the latter are dependent on nuclear models, different methods give qualitatively consistent matter radii, as indicated by the gray bands in the figure. In both $^6$He and $^8$He, the matter radii are significantly larger than the point-proton radii, a clear signature of the core-halo structure of these nuclei. The matter radius for $^4$He should be the same as the indicated point-proton radius. Also given in Fig. 8 are the values from *ab initio* calculations based on the no-core shell model (NCSM) (Maris, 2013) and Green's function Monte Carlo (GFMC) techniques. The nuclear models and their calculations of the radii are discussed in the following section.

It should be noted that the experimental point-proton radii of $^4$He, $^6$He and $^8$He would be larger by 0.021 fm, 0.016 fm and 0.017 fm, respectively, were the PDG value of the proton charge radius replaced with that obtained from the recent muonic hydrogen studies (see Sec II.B). This change would be significant compared to the experimental uncertainties; however, it would not alter the conclusion on the nuclear structure of these isotopes that are predominantly based on relative changes in the charge radii along the isotope chain.

## VI. MATTER RADII OF $^{6,8}$HE

As is discussed in previous sections, charge radii are determined by the precisely known electromagnetic interactions and, hence, have no dependence on models of nuclear structure or hadronic interaction. Point-proton radii can be derived from the charge radii via Eq. (4) with weak dependence in the correction terms on nuclear models. On the other hand, point-neutron radii or matter radii can only be probed with hadronic interactions whose quantitative understanding depends strongly on nuclear models. Furthermore, such probes can significantly modify the target nucleus as they interact with it. Thus, the extraction of matter radii is subject to significant model dependence in the analysis of the experiments, particularly in nuclear reaction measurements.

Although the short-lived isotopes are thus far too scarce to form a target, it has been possible since the mid-1980s to have these isotopes produced and selected in flight to form a secondary beam for scattering experiments. Such measurements are now regularly performed at the world's premier radioactive isotope facilities to study nuclear reactions, nuclear structure, and to determine nuclear properties including radii.

### A. Nuclear reaction measurements

Tanihata *et al.* (Tanihata *et al.*, 1985b) first demonstrated the use of radioactive isotope beams to measure interaction cross sections, defined as total nuclear reaction cross sections, based on which the radii of the He isotopes ($^{3,4,6,8}$He) were deduced. The radii of $^{6}$He and $^{8}$He were found to increase from $^{4}$He faster than the canonical $A^{1/3}$ rule, thus providing the first observational support for the neutron halo picture. This method has also been used to study other exotic isotopes including Li, Be, B, and C (Tanihata *et al.*, 1985a, 1988).

A series of experiments was performed at the Bevalac accelerator facility of the Lawrence Berkeley National Laboratory. Secondary beams of He isotopes were produced through projectile fragmentation of an 800 MeV/u $^{11}$B primary beam, separated by a magnetic analyzer, and directed onto a target made of either Be, C, or Al. The transmission of the He isotopes through the various targets was measured using a spectrometer-detector assembly. The interaction cross section was determined with a relative accuracy of 1%.

The matter radius can be extracted from the interaction cross section using a semiclassical optical model (Karol, 1975) based on both the free nucleon-nucleon collision cross section and a model of the spatial distribution of nucleons inside the nucleus. The top three rows of Table VI list matter radii extracted from the same set of interaction cross sections under three different models of the nucleon distribution. The extracted matter radii of $^{4}$He and $^{8}$He appear to be insensitive to the models. On the other hand, the extracted matter radius of $^{6}$He ranges from 2.33 fm to 2.71 fm, perhaps due to its weak binding energy and halo nature (Al-Khalili *et al.*, 1996; Tanihata *et al.*, 1988, 1992). The measurements do not distinguish between a proton and a neutron. When it is assumed that a $^{6}$He or $^{8}$He nucleus is formed by an undistorted $^{4}$He-like core and valence neutrons, both the point-proton radius and the point-neutron radius can be deduced. In the case of $^{6}$He, the extracted value of its point-proton radius ranges from 1.72 fm to 2.21 fm (Tanihata *et al.*, 1988, 1992). However, the point-proton radius of $^{4}$He extracted from the precisely measured charge radius using Eq. (4) is $r_p = r_m = 1.462$ fm, which is much smaller than the Tanihata values, but in good agreement with the analysis of Alkhzov *et al.* (see Table VI).

TABLE VI Experimental and theoretical matter radii of $^{4,6,8}$He. In the top three rows, the values were extracted from the same set of interaction cross sections but with nucleon distribution functions derived from three different limits of the Glauber model. The values in the 4th row were extracted from elastic scattering data. The ones in the 5th were derived from the same elastic scattering data, but with a distribution model that included a long tail of neutrons on the outside. Theoretical matter radii extracted using Green's function Monte Carlo (GFMC) and no-core shell model (NCSM) methods are listed in the last two rows. Units are fm.

| | Matter radii (fm) | | | |
|---|---|---|---|---|
| Method | $^{4}$He | $^{6}$He | $^{8}$He | Ref. |
| Interaction | 1.57(4) | 2.48(3) | 2.52(3) | a) |
| Interaction | 1.63(3) | 2.33(4) | 2.49(4) | b) |
| Interaction | 1.58(4) | 2.71(4) | | c) |
| Elastic | 1.49(3) | 2.30(7) | 2.45(7) | d) |
| Elastic, tail | | 2.45(10) | 2.53(8) | d) |
| GFMC, AV18+IL7 | 1.435(3) | 2.58(7) | 2.55(4) | e) |
| NCSM, JISP16 | 1.45(1) | 2.38(7) | >2.5 | f) |

a) (Tanihata *et al.*, 1988), b) (Tanihata *et al.*, 1992), c) (Al-Khalili *et al.*, 1996), d) (Alkhazov *et al.*, 2002), e) This work, f) (Maris, 2013)

### B. Elastic scattering measurements

At the SIS heavy-ion synchrotron of GSI, secondary beams of $^{4,6,8}$He with an energy of approximately 0.7 GeV/u were produced by fragmentation, isotopically selected by the fragment separator FRS, and were then incident upon a hydrogen-filled target chamber. Differential cross sections for p–$^{6}$He and p–$^{8}$He elastic scattering were measured in inverse kinematics at small momentum transfers up to 0.05 GeV$^2$, based on which nuclear radii were extracted with a relative precision of 3% (Alkhazov *et al.*, 1997, 2002). This versatile method has also been applied to the exotic Li (Dobrovolsky *et al.*, 2006) and Be isotopes (Ilieva *et al.*, 2012).

The basic principle of this method is similar to that of $eA$ collisions: the slope of the differential cross section vs. $q^2$ near $q^2 = 0$ is related simply to the matter radius. In practice, this method, like the interaction cross section one, depends on nuclear models. For example, it requires detailed knowledge of the scattering amplitudes of proton-proton collisions, which are well determined, and proton-neutron collisions, where data are scarce. Furthermore, due to the short-range nature of the interaction potential among nucleons, the radii also depend on the assumed model of the spatial distribution of nucleons inside the nucleus. Assuming that $^{6}$He (or $^{8}$He) is formed by a $^{4}$He-like core and a neutron halo, Alkhazov *et al.* (Alkhazov *et al.*, 2002) showed that the matter radius changes little between a halo of a Gaussian distribution and of a $1p$ harmonic oscillator distribution, even though these two distributions differ significantly in their values at the origin. The resulting matter radii are listed in the 4th row of Table VI. Under these assumptions, the core radii are equivalent to the point-proton radii, and were derived to be 1.88(12) fm for $^6$He and 1.55(15) fm for $^8$He. On the other hand, for $^6$He, a halo distribution with an assumed long tail causes the matter radius to shift up by 0.15 fm, or 6% (5th row of Table VI).

## VII. *AB INITIO* CALCULATIONS OF $^{4,6,8}$HE RADII

The goal of *ab initio* calculations of light nuclei is to understand these systems as collections of nucleons interacting via realistic interactions through solutions of the many-nucleon Schrödinger equation. In this review, those methods that ignore the nucleon structure of the $^4$He core or that do not use realistic forces (forces that reproduce observed two-nucleon scattering data) are not considered. There are two main challenges in microscopic few- and many-nucleon calculations: (1) determining the proper Hamiltonian, and (2) given a Hamiltonian, accurately solving the Schrödinger equation for $A$ nucleons. The Hamiltonian is discussed in the next subsection and two *ab initio* methods are presented next. Finally, there is a more qualitative discussion of the nuclear density distributions.

### A. The Hamiltonian

QCD has been firmly established as the fundamental theory of the strong interaction. However, at the low energy scale comparable to the binding energy of the nucleus, the theory is nonperturbative and still cannot be used to calculate the interacting potential between nucleons. Instead, the Hamiltonian is developed based on effective nuclear potential models of the form

$$H = \sum_i K_i + \sum_{i<j} v_{ij} + \sum_{i<j<k} V_{ijk} \tag{13}$$

Here $K_i$ is the nonrelativistic kinetic energy, taking into account the mass difference between the neutron and proton, $v_{ij}$ is the nucleon-nucleon (*NN*) potential, and $V_{ijk}$ is the three-nucleon *NNN* potential.

In the 1990's, a number of *NN* potentials were formulated based on meson-exchange principles. Among them, Argonne V18 (Wiringa *et al.*, 1995) (AV18) is the most commonly used. A local potential written in operator format, AV18 contains a complete representation of the pp, pn and nn electromagnetic terms, the long-ranged one-pion exchange terms, and phenomenological shorter-ranged terms; the strong-interaction terms are expressed as 18 local spin-isospin operators. The parameters of these terms are determined by fitting the large body of *NN* scattering data.

It has long been known that calculations with just *NN* potentials fail to reproduce the binding energies of nuclei; three-nucleon (*NNN*) potentials are also required. (The JISP potentials described below are a special exception.) These arise naturally from an underlying meson-exchange picture of the nuclear forces or from chiral effective field theories. Unfortunately, much *NNN* scattering data are well reproduced by calculations using just *NN* forces, so the *NNN* forces have to be determined from properties of light nuclei. Illinois-7 (Pieper, 2008a) (IL7) is the latest of the series of the Illinois three-body potentials (Pieper *et al.*, 2001) that were developed for use with AV18. It consists of two- and three-pion terms and simple phenomenological repulsive terms. The two-pion term contains the well-known Fujita-Miyazawa term (Fujita and Miyazawa, 1957) present in all realistic *NNN* potentials. In it, a pion is exchanged between two nucleons, exciting one of them to a resonance. The resonance then decays back to a nucleon by emitting a pion to the third nucleon. This is the longest-range *NNN* potential and is attractive in all nuclei and in nuclear matter.

Nucleon-nucleon scattering determines only the on-shell properties of the *NN* potential. This is sufficient to completely specify the potential if one requires it to be local, as is the case with AV18. However, many choices of nonlocal behavior can be made. This freedom has been used to construct *NN* potentials that give correct binding energies of nuclei without an additional *NNN* potential (Shirokov *et al.*, 2007). Inverse scattering methods are used to construct a potential with parameters that can be fit to nuclear binding energies without changing the fit to *NN* phase shifts. The most recent version of these potentials is JISP16, which gives a good reproduction of energies of light nuclei (Maris *et al.*, 2009).

More recently, systematic expansions based on chiral effective field theory, $\chi$EFT, have been developed (Epelbaum, 2006; Machleidt and Entem, 2011). An up-to-date review of this subject, including a discussion of the three-nucleon potentials, is provided by Epelbaum and Meissner (Epelbaum and Meissner, 2012). The expansion is carried out to the third order: N$^3$LO. Like in AV18, these potentials also contain a large number of parameters that must be determined by fits to *NN* scattering data. Three-nucleon potentials generated systematically for the $\chi$EFT potentials are now available up to only second order; these are used with the third-order *NN* potential. They contain two parameters that must be determined from fits to properties of nuclei. The authors are not aware of calculations of $^{6,8}$He radii using these potentials.

### B. Green's function Monte Carlo (GFMC)

Once the Hamiltonian is given, many-nucleon calculations can be carried out using the Green's function Monte Carlo (GFMC) method. (GFMC is often referred to as diffusion Monte Carlo in other areas of physics.) Heuristic introductions to the nuclear Variational Monte Carlo (VMC) and GFMC methods are given in (Pieper, 2008b;

Pieper and Wiringa, 2001) and detailed descriptions are in (Pieper *et al.*, 2004) and references therein.

For the first step of the calculation, in VMC, the parameters of a trial wave function, $\Psi_T$, are varied to minimize the expectation value of $H$,

$$E_T = \frac{\langle \Psi_T | H | \Psi_T \rangle}{\langle \Psi_T | \Psi_T \rangle} \geq E_0 \qquad (14)$$

The resulting energy $E_T$ is, by the Raleigh-Ritz variational principle, greater than or equal to the true ground-state energy for the quantum numbers ($J^\pi$, $J_z$, $T$, and $T_z$) of $\Psi_T$.

The wavefunction $\Psi_T$ contains one-, two-, and three-body correlations. The radial parts of these terms can be chosen at will; they are tabulated on a radial grid as opposed to being expanded in a basis set. This is important for the weakly-bound helium isotopes; the one-body correlations are bound-state wavefunctions that are computed as solutions of the Woods-Saxon wells and have long tails. The two-body correlations are of the Jastrow form (a product over all nucleon pairs) containing both central and noncentral operators. The central part is small at short distances and serves to keep the nucleon pairs from seeing the strong repulsive core in $v_{ij}$. The noncentral part contains the important operators of $v_{ij}$, in particular the tensor correlation. Finally the three-body correlation also contains important operators from $V_{ijk}$. Due to the complicated structure of $\Psi_T$, the $3A$ dimensional integral in $\langle \Psi_T | H | \Psi_T \rangle$ cannot be factorized. It is evaluated by the Metropolis Monte Carlo method (Metropolis *et al.*, 1953).

Despite their complexity, the VMC trial wave functions are not good enough for $p$-shell nuclei. They contain admixtures of excited state components in addition to the desired exact ground state component $\Psi_0$. To overcome this problem, GFMC is used to project $\Psi_0$ out of $\Psi_T$ by propagating in imaginary time, $\tau$:

$$\begin{aligned} \Psi(\tau) &= \exp[-(H' - \tilde{E}_0)\tau]\Psi_T \\ \lim_{\tau\to\infty} \Psi(\tau) &\propto \Psi_0 \end{aligned} \qquad (15)$$

where $H'$ is an approximation to the desired Hamiltonian $H$, and $\tilde{E}_0$ is a guess for the exact energy $E_0$. The evaluation of $\exp[-(H' - \tilde{E}_0)\tau]$ is made by a sequence of small steps $\triangle\tau$ in imaginary time using an approximation to $\exp(-H\triangle\tau)$. Each step involves a full $3A$-dimensional integral done by Monte Carlo. These steps are made until there are only statistical fluctuations in the energy at each step. They are then continued until an average over these statistically fluctuating values has the desired precision.

GFMC calculations using AV18 with the Illinois $V_{ijk}$ are successful in reproducing the energies of nuclear states for $A \leq 12$ (Pieper, 2005, 2008b; Pieper *et al.*, 2004). The He isotope energies and corresponding two-neutron separation energies $E_{2n}$ obtained for AV18+IL7 are given in Table V. Quantities other than energy may converge at a much slower pace. This is particularly true for the radii of weakly bound states. For $^8$He, and especially for $^6$He, there are long-term fluctuations in the radii as GFMC calculations propagate in imaginary time. These fluctuations are associated with the small two-neutron separation energies; according to the calculations, $E_{\rm sep} = 0.97$ MeV for $^6$He and 1.86 MeV for $^8$He. Fig. 9 displays the results of multiple GFMC calculations with different initial conditions. Even though GFMC can compute the binding energies precisely, its relative errors in $E_{\rm sep}$ are significant since, for $^6$He, $E_{\rm sep}$ is only 3% of the binding energy. For example, changes in the starting wavefunction $\Psi_T$ and other aspects of the GFMC calculations can result in changes of 0.2 MeV in $E_{\rm sep}$, or a few percent change in the radius. For these weakly bound nuclei, more precise values of radii can be obtained by selecting those calculations that simultaneously yield the experimentally known $E_{\rm sep}$ value, marked with a star in Fig. 9, with its associated range interpreted as an uncertainty of the computed radii. The same procedure is used to get the matter radii $r_m$ (Table VI). The computed point-proton radii $r_p$ for all three helium isotopes are in excellent agreement with the experimental values (Fig. 8). The computed matter radii $r_m$ are in reasonable agreement with the results of nuclear scattering experiments.

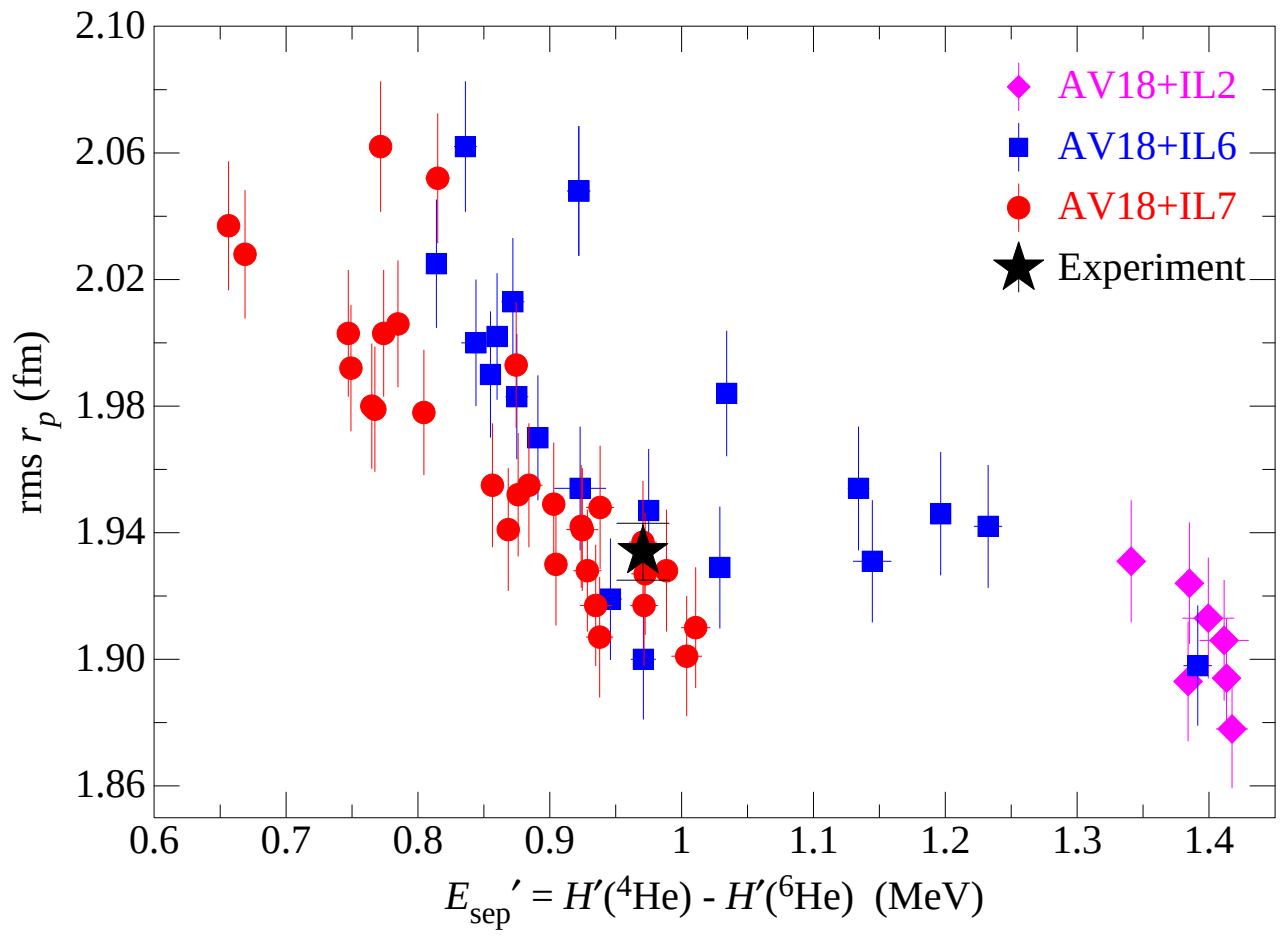

FIG. 9 $^6$He point-proton radius vs. two-neutron separation energy obtained from a number of GFMC calculations with different initial conditions and three-body potentials.

Figure 10 presents the point-proton and point-neutron densities of $^{4,6,8}$He extracted from GFMC calculations that give $E_{\rm sep}$ close to the experimental values. $^4$He is extremely compact. Its central density is twice that of nuclear matter, and it has essentially identical proton and neutron densities. Fig. 10 clearly indicates that $^{6,8}$He have large neutron halos due to the weak binding of the extra neutrons. The neutron halo of $^6$He is more diffuse than that of $^8$He as is expected from the smaller $E_{\rm sep}$ of $^6$He.

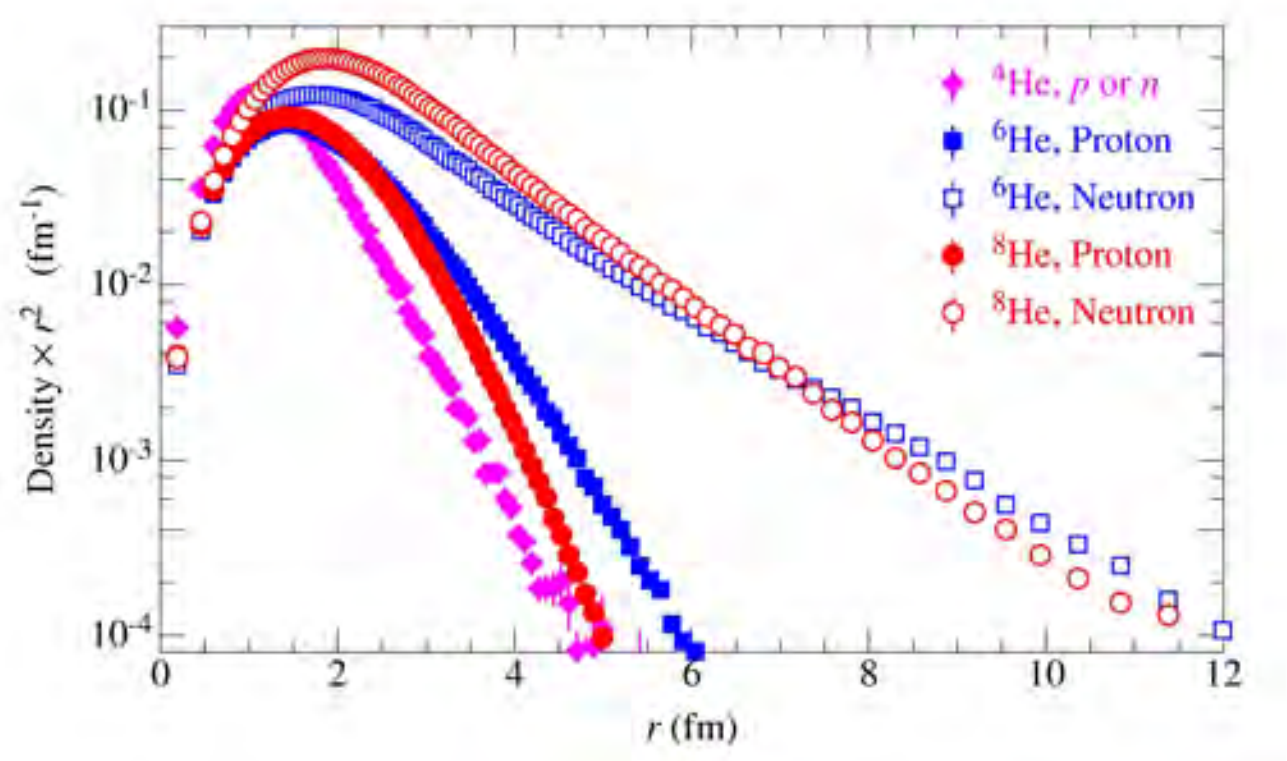

FIG. 10 Point-proton and point-neutron densities of the even helium isotopes as extracted from GFMC calculations.

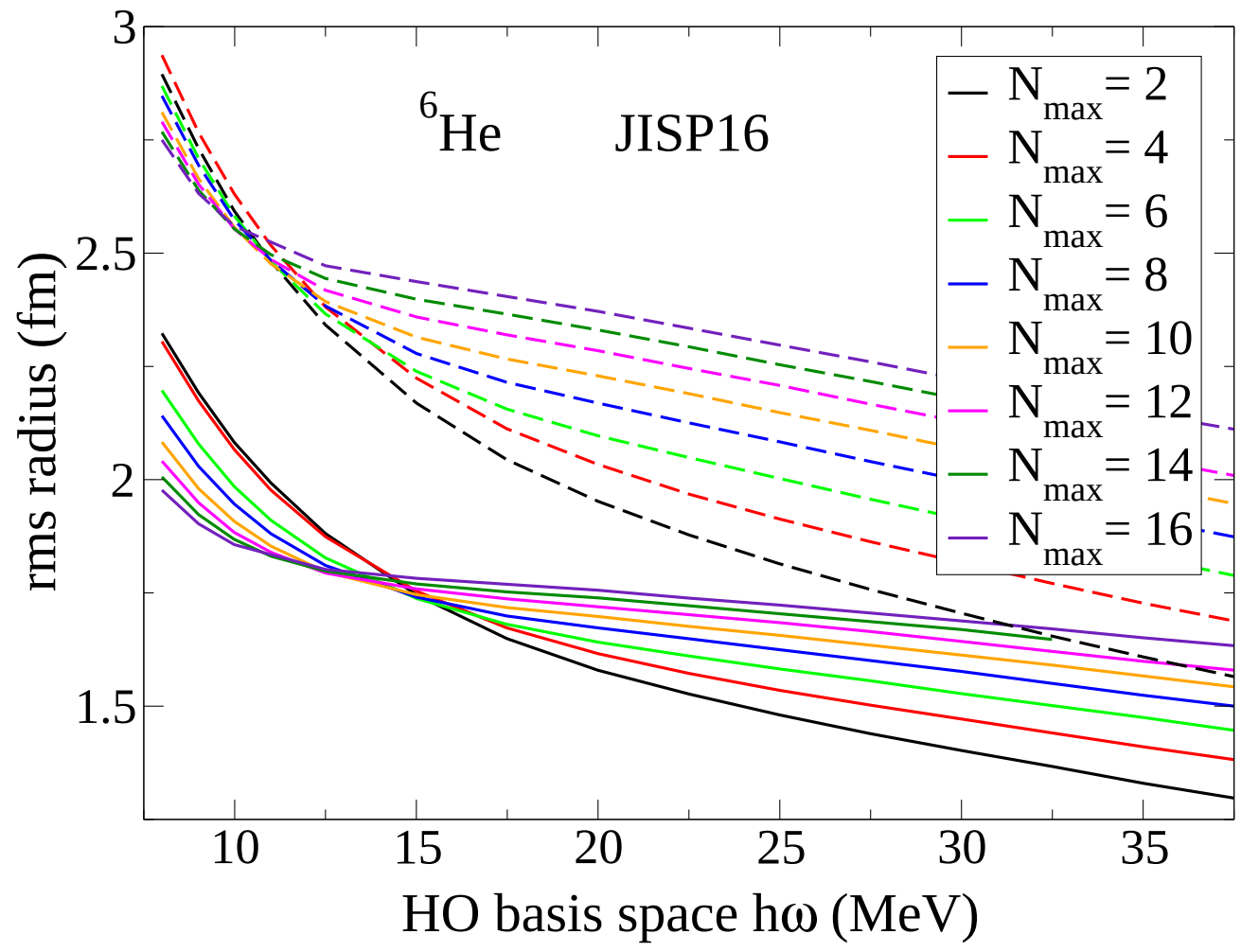

FIG. 11 Model-space dependence of the $^{6}$He point proton (solid curves) and neutron (dashed curves) radii in NCSM calculations using the JISP16 *NN* potential. Figure from (Maris, 2013).

### C. No-Core Shell Model (NCSM)

The *ab initio* no-core shell model (NCSM) is another approach for solving the nuclear many-body problem for light nuclei (Barrett *et al.*, 2013; Navrátil *et al.*, 2000). In the traditional shell model, only the valence nucleons outside an inert core are treated. NCSM employs many techniques developed for the traditional shell model calculations, but treats all nucleons as active. In NCSM, a nucleus is considered as a system of $A$ point-like non-relativisitic nucleons that interact via realistic *NN* and *NNN* potentials. The operators and wave functions are expressed in a finite space of a harmonic oscillator (HO) basis, truncated by a judiciously chosen maximal HO excitation energy that defines the size of the model space. In order to account for short-range correlations and to speed up convergence, an effective interaction potential is constructed from the original potentials by means of a unitary transformation. The results depend upon both the frequency $\omega$ and number of shells $N_{\rm max}$ of the HO basis. As $N_{\rm max}$ is increased, the dependence on $\omega$ is reduced and converged results are obtained. Extrapolation of $N_{\rm max}$ to infinity is sometimes necessary. The HO basis has the wrong asymptotic behavior for bound-state wave functions. This can cause particular difficulties for extracting rms radii, as is discussed in Ref. (Cockrell *et al.*, 2012).

The ground state properties of $^{4}$He, $^{6}$He and $^{8}$He were calculated with NCSM using the JISP16 *NN* potential (Maris, 2013). The calculations for $^{6}$He and $^{8}$He were performed in model spaces up to $N_{\rm max} = 16$ and $N_{\rm max} = 14$, respectively, for a wide range of HO frequencies. Figure 11 shows the HO frequency dependence of the values of the $^{6}$He point-proton and point-neutron radii for different HO $N_{\rm max}$. A general feature is a decrease of the HO frequency dependence with increasing $N_{\rm max}$. However the extrapolation to $N_{\rm max} = \infty$ is still substantial, and only a lower limit can be placed on the point-neutron radius of $^{8}$He. The extrapolated results are summarized in Tables V and VI and Fig. 8. The $^{6}$He values of $E_{2n} = 0.50(5)$ and $r_p = 1.85(5)$ are considerably off the trend of the GFMC values of Fig. 9.

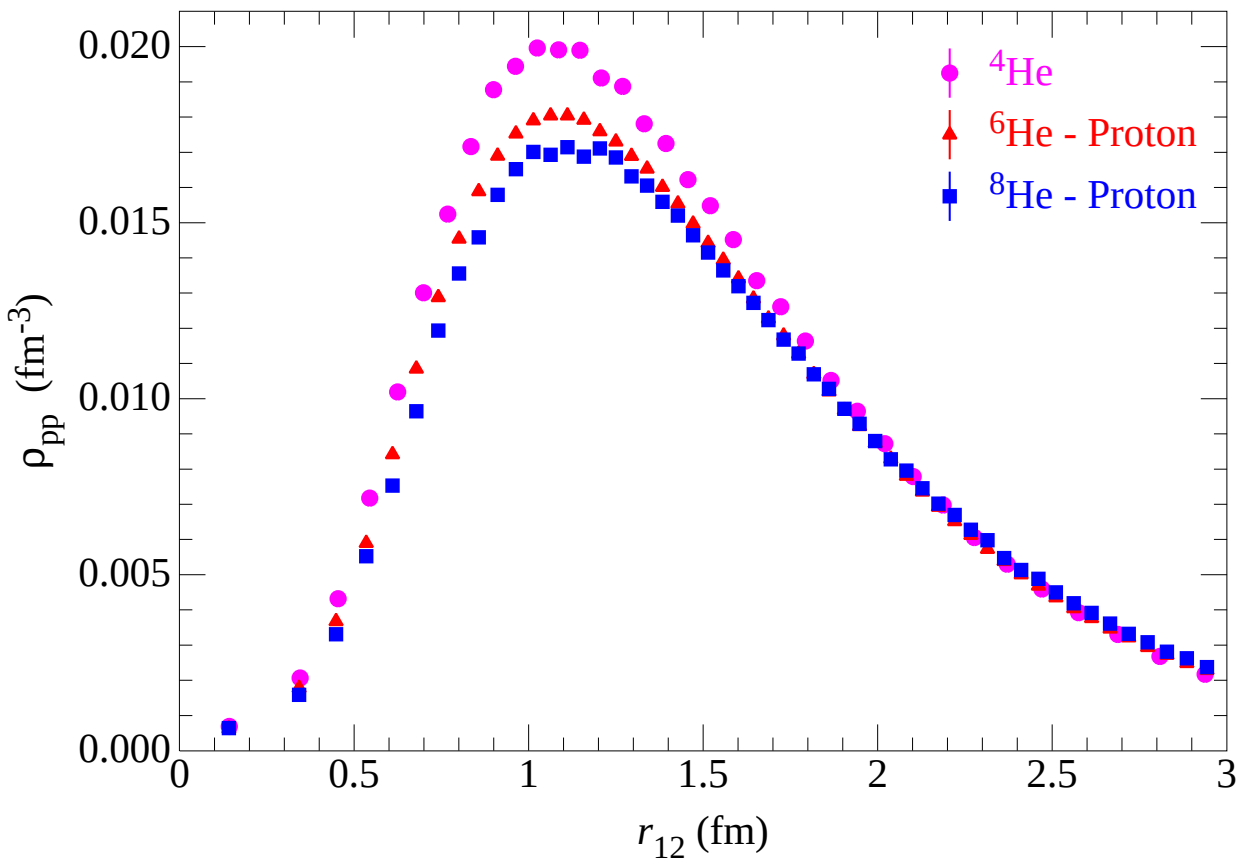

FIG. 12 Proton-proton pair densities $\rho_{pp}$ of the even helium isotopes as extracted from GFMC calculations. From (Pieper, 2008a)

### D. Contributions to the charge radii of $^{6,8}$He

The proton distribution in $^{6,8}$He is much more spread out than the density of $^{4}$He (see Sec. VII.B), even though $^{6,8}$He have only extra neutrons added to the $^{4}$He core. This might be thought to indicate the so-called "core swelling" effect — the core of $^{6,8}$He is enlarged by the presence of the valence neutrons. This effect can be stud-ied by computing $\rho_{pp}$, the pair density which is proportional to the probability for finding two protons a given distance apart (Pieper, 2008a) (Fig. 12). Unlike the one-body densities, these distributions are not sensitive to center-of-mass effects. GFMC calculations indicate that the $pp$ distribution spreads out only slightly with neutron number in the helium isotopes, with an increase of the pair rms radius of approximately 4% in going from $^4$He to $^6$He, and 8% from $^4$He to $^8$He. While this could be interpreted as a swelling of the $\alpha$ core, it might also be due to charge-exchange correlations induced by terms in $v_{ij}$ which exchange protons and neutrons and, thus, transfer charge from the core to the valence nucleons. Since these correlations are rather long-range, they can have a significant effect on the $pp$ distribution. VMC calculations of $^4$He with wave functions modified to give $\rho_{pp}$ distributions close to those of $^{6,8}$He suggest that the $\alpha$ cores of $^{6,8}$He are excited by ∼80 and ∼350 keV, respectively, which corresponds to only a 0.4−2% admixture of the first $0^+$ excited state of $^4$He at 20 MeV. Thus, the observed increase in the rms radius of the proton density is largely due to the $\alpha$ core of $^{6,8}$He being "pushed around" by the neutrons in the so-called recoil effect.

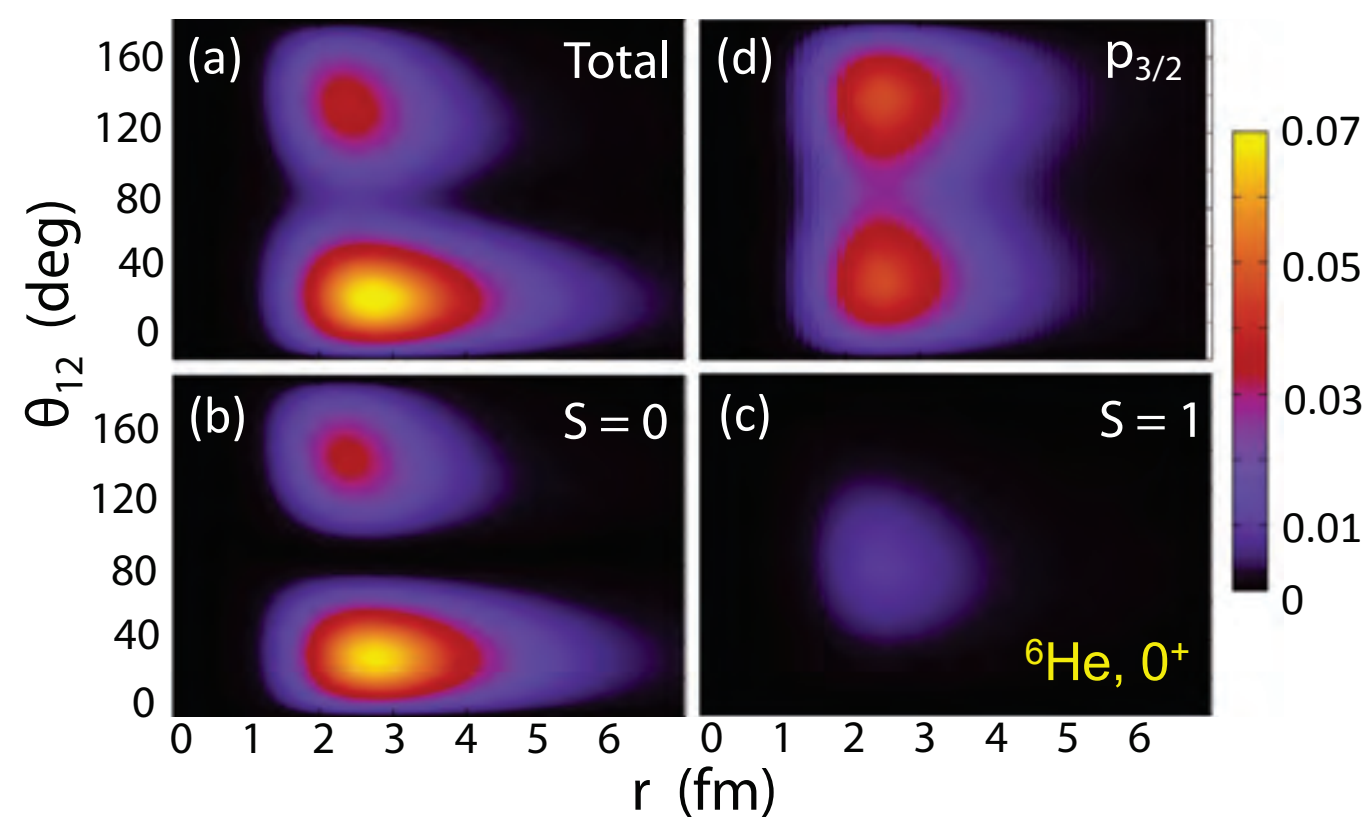

FIG. 13 Density distributions of valence neutrons in $^6$He for the special case of both neutrons being equally distant from the $^4$He core. The densities are shown as a function of the distance of the neutrons from the core and the angle between them relative to the $^4$He core. From (Papadimitriou *et al.*, 2011)

The correlation of valence neutrons has been studied within the complex-energy configuration-interaction framework (Hagino and Sagawa, 2005; Kikuchi *et al.*, 2010; Papadimitriou *et al.*, 2011) (Fig. 13). The density distribution of the valence neutrons exhibits two maxima: (1) a dineutron configuration that corresponds to a small opening angle and a large radial extension; and (2) a cigar-like configuration that corresponds to a radially localized region with large angles. These authors find that the reduction of the charge radius from $^6$He to $^8$He is not due to a more even distribution of valence neutrons around the core, but rather to the reduction of the amplitude of the dineutron configuration in the ground-state wavefuction, resulting in a smaller core recoil radius.

TABLE VII Parameters for the determination of nuclear radii from the measured isotope shifts in three-electron atomic cases: lithium atom and beryllium ion (see Eq. 12). The field shift constant K includes a relativistic correction (Puchalski *et al.*, 2006). A nuclear polarization correction $\Delta E_{\rm pol}$ of 39(4) kHz is included for $^{11}$Li (Puchalski *et al.*, 2006) and of 208(21) kHz for $^{11}$Be (Puchalski and Pachucki, 2008). It is negligible for most other isotopes, but a finite contribution of < 60 kHz is estimated for $^{12}$Be (Krieger *et al.*, 2012).

| Isotopes | Transition | $\delta\nu_{\rm MS}$ (MHz) | $K$ (MHz/fm$^2$) |
|---|---|---|---|
| $^7$Li – $^6$Li | $2\,^2S_{\frac12} - 3\,^2S_{\frac12}$ | –11452.8211(28) | 1.572 |
| $^8$Li – $^6$Li | $2\,^2S_{\frac12} - 3\,^2S_{\frac12}$ | –20087.8026(50) | 1.572 |
| $^9$Li – $^6$Li | $2\,^2S_{\frac12} - 3\,^2S_{\frac12}$ | –26784.6213(67) | 1.572 |
| $^{11}$Li – $^6$Li | $2\,^2S_{\frac12} - 3\,^2S_{\frac12}$ | –36554.325(9) | 1.570 |
| $^7$Be – $^9$Be | $2\,^2S_{\frac12} - 2\,^2P_{\frac12}$ | –49225.779(38) | 17.02 |
| $^{10}$Be – $^9$Be | $2\,^2S_{\frac12} - 2\,^2P_{\frac12}$ | 17310.441(12) | 17.03 |
| $^{11}$Be – $^9$Be | $2\,^2S_{\frac12} - 2\,^2P_{\frac12}$ | 31560.294(24) | 17.02 |
| $^{12}$Be – $^9$Be | $2\,^2S_{\frac12} - 2\,^2P_{\frac12}$ | 43390.168(39) | 17.02 |

## VIII. THREE-ELECTRON ATOMS: LI AND BE$^+$

Just as for helium (see Eq. (9)), the precision theory of a three-electron system, such as a neutral atom of lithium (Li) or singly-charged ion of beryllium (Be$^+$), requires fully correlated calculations in Hylleraas coordinates, but involving now all six interparticle distances, $r_1$, $r_2$, $r_3$, $r_{12}$, $r_{23}$, and $r_{31}$. This makes the calculations much more difficult because the number of terms in the basis set grows very rapidly with the highest powers included, and the nine-dimensional integrals are much more difficult to calculate analytically. Recent advances have now achieved the level of accuracy needed for both the nonrelativistic energies and the relativistic corrections (Krieger *et al.*, 2012; Nörtershäuser *et al.*, 2011b; Puchalski and Pachucki, 2008; Yan and Drake, 2000; Yan *et al.*, 2008). Of particular importance and difficulty are the three-electron Bethe logarithms needed to calculate the QED energy shift and its mass dependence (Puchalski *et al.*, 2013; Yan *et al.*, 2008). As a result, atomic spectroscopic measurements on an isotope of Li or Be$^+$ can be used to deduce its corresponding nuclear charge radius and nuclear moments (Table VII). Extensive studies have since been carried out on the neutron-rich Li and Be$^+$ isotopes, revealing halo and cluster formation in these nuclei.

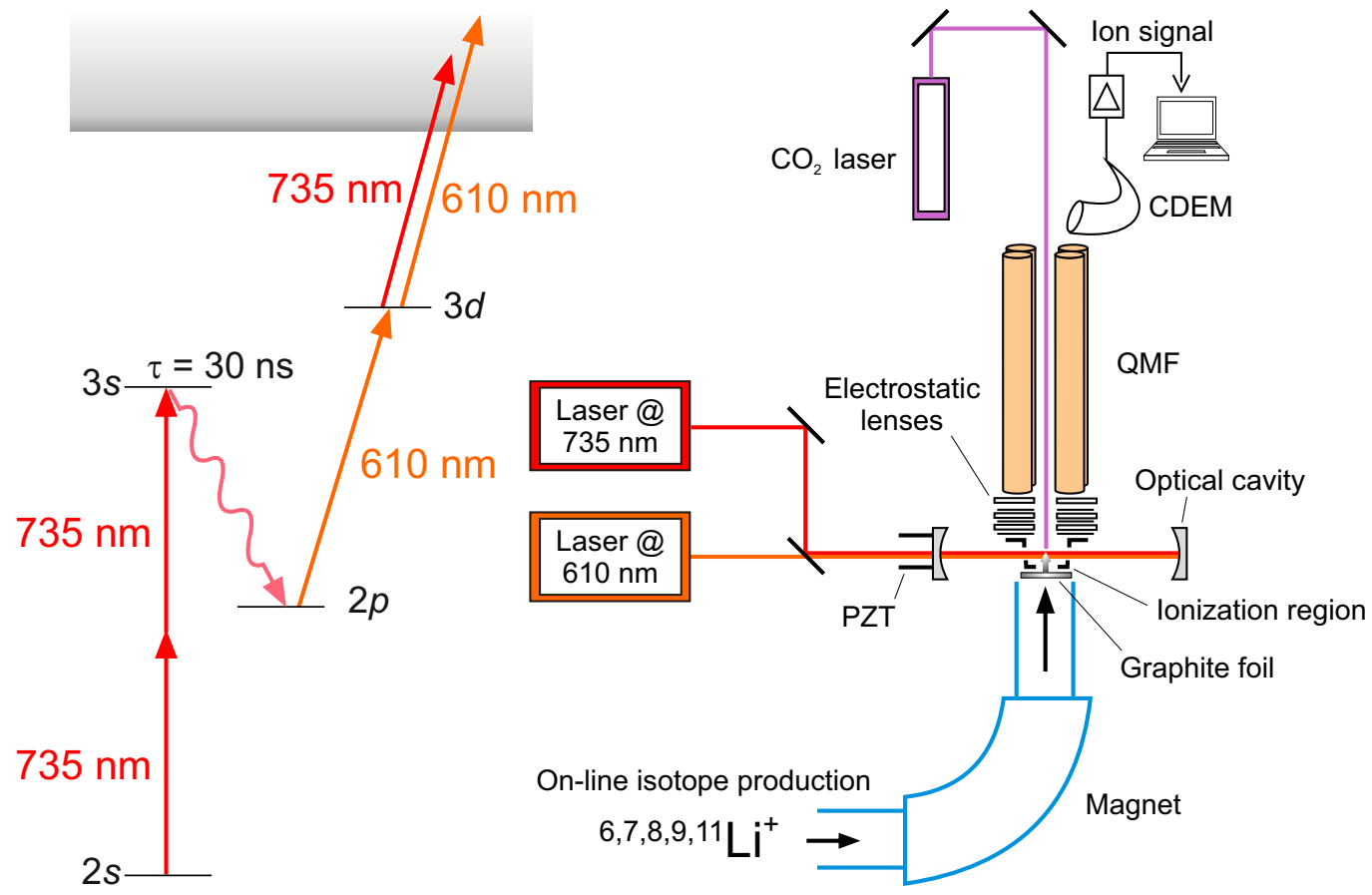

FIG. 14 (Color online) Laser excitation schme (left) and experimental setup (right) to measure the $2s\ ^2S_{1/2} \to 3s\ ^2S_{1/2}$ electronic transition in lithium. CDEM, continuous dynode electron multiplier; QMF, quadrupole mass filter; PZT, piezoelectric transducer. Modified figure from (Nörtershäuser *et al.*, 2011b).

### A. Laser spectroscopic studies of $^{6,7,8,9,11}$Li

Laser spectroscopy has been performed on every lithium isotope with a bound nucleus; i.e. $^{6,7,8,9,11}$Li (Bushaw *et al.*, 2007; Nörtershäuser *et al.*, 2011a,b; Sánchez *et al.*, 2009). $^8$Li and $^9$Li were produced with moderate yields using a $^{12}$C beam of 11.4 MeV/u at GSI (Ewald *et al.*, 2004), and with high yields using fragmentation in a tantalum target induced by a 500 MeV proton beam at TRIUMF. In both cases the reaction products were surface ionized, mass separated in a sector magnet and transported to the experimental setup as an ion beam at a a beam energy of about 40 keV. The reaction used at TRIUMF provided also the halo isotope $^{11}$Li at a rate of 30 000 s$^{-1}$ (Sánchez *et al.*, 2006). This isotope is demanding because of its very short half-life of only 8.4 ms. Therefore, the spectroscopic technique had to provide a high efficiency while being fast and highly accurate. This was achieved combining a Doppler-free two-photon transition in the atomic system for high-resolution spectroscopy with an efficient detection by means of resonance ionization mass spectrometry (RIMS) (Fig. 14).

The spectroscopic approach required the short-lived species to be prepared as neutral atoms at low-energy. Thus, the lithium ions were stopped in a graphite foil of approximately 300 nm thickness and quickly boiled out as neutral atoms with little loss by heating the foil to a temperature of 2000 K with a $CO_2$-laser beam of approximately 4 W. After drifting a few mm through the entrance hole of a quadrupole mass filter's (QMF) ionization region, being at a slightly positive potential to suppress surface ions, the thermal atoms cross the superimposed beams of a titanium:sapphire laser at 735 nm and a dye laser at 610 nm. Doppler-free, two-photon excitation of the $2s - 3s$ transition was induced with the 735-nm beam. The excited atoms decay into the $2p\ ^2P_{1/2,3/2}$ manifold from which they are resonantly ionized through the $3d\ ^2D_{3/2,5/2}$ levels as shown in Fig. 14. The non-linear two-photon excitation as well as the non-resonant ionization require relatively high beam intensities. Therefore, both laser beams were resonantly enhanced by a factor of 50 – 100 in a two-color optical cavity. The ions produced inside the ionization region are extracted with the QMF's ion optics, mass filtered and detected individually by a continuous dynode electron multiplier. Recording the number of ionized atoms as a function of the Ti:Sa laser frequency, tuned typically across a 200 – 400 MHz region, both hyperfine components of the $2s - 3s$ transition were detected and fitted with an appropriate line profile. The isotope shifts for this transition, relative to $^6$Li, were precisely determined. The AC-Stark shift contribution was considered by extrapolating back to zero laser intensity and the possibility of isotope-dependent contributions to this shift was carefully investigated. With reference to the charge radius of $^6$Li — a value independently determined by electron scattering experiments — the nuclear charge radii of all the other Li isotopes were deduced (Table VIII and Fig 15).

The results were compared to the predictions of a number of nuclear structure models (Nörtershäuser *et al.*, 2011a). Among those, cluster models showed the best agreement with the experimental charge radii, but the comparison with the experimentally determined nuclear moments is often not as convincing (Neugart *et al.*, 2008). While many microscopic models are able to reproduce the trend of nuclear charge radii from $^6$Li to $^9$Li, they face a tough challenge in the description of $^{11}$Li. Table VIII and Fig. 15 provide the results of the GFMC calculations of $r_p$ with the same AV18+IL7 Hamiltonian that was also used for the helium isotopes. At present, these calculations have been done only up to $A = 10$. The agreement with experiment is good for $^{6,7}$Li, but then the theoretical values fall too rapidly. It should be noted that a combination of the nuclear charge radii, matter radii, and information from Coulomb dissociation provides in principle sufficient information to separate the contributions from center-of-mass motion and from intrinsic core excitations (Esbensen *et al.*, 2007; Nörtershäuser *et al.*, 2011a). Unfortunately, the accuracy of the available data and the remaining model dependence still do not allow for an unambiguous separation.

The difference between the charge radii of the stable $^6$Li and $^7$Li has already been determined for four transitions in two charge states as summarized in (Nörtershäuser *et al.*, 2011b). While the $1s2s\,^3S_1 - 1s2p\,^3P_2$ transition in Li$^+$ and the $2s - 3s$ two-photon transition in neutral Li lead to consistent values of $\delta\left\langle r_c^2\right\rangle$, inconsistent values were reported for the D1 and D2 lines in neutral Li. This inconsistency was recently resolved by including the contribution of the quantum interference effects to the lineshapes, particularly of the D2 transition with unresolved hyperfine structure (Sansonetti *et al.*, 2011). Taking this into account, the charge radii determined from different transitions are all in agreement within the reported uncertainties (Brown *et al.*, 2013). Moreover, the splitting isotope shift between the D1 and the D2 line is now also in agreement with the theoretical prediction, corroborating the internal consistency of the calculations.

TABLE VIII Half-lives, spin-parities, experimental charge radii and experimental and GFMC point-proton radii in the Li and Be isotope chains. Charge radii are based on isotope shift measurements in Li atoms (Nörtershäuser *et al.*, 2011a) and $Be^+$ ions (Krieger *et al.*, 2012) and are referenced to the values of the stable $^6$Li and $^9$Be, respectively, which are independently determined from electron scattering experiments. The first uncertainty of the charge radii has been determined from the quoted error of the isotope shift, whereas the second one includes the uncertainty of the reference radius. Radii are in fm. The GFMC values (Pastore *et al.*, 2012) are for the AV18+IL7 Hamiltonian.

| Isotope | $t_{1/2}$ | $J^\pi$ | $r_c$ | $r_p$ | |
|---|---|---|---|---|---|
| | | | | Expt. | GFMC |
| $^6$Li | *stable* | $1^+$ | 2.589(0)(39) | 2.45(4) | 2.39(1) |
| $^7$Li | *stable* | $3/2^-$ | 2.444(4)(43) | 2.31(5) | 2.28(1) |
| $^8$Li | 840 ms | $2^+$ | 2.339(7)(45) | 2.20(5) | 2.10(1) |
| $^9$Li | 180 ms | $3/2^-$ | 2.245(7)(47) | 2.11(5) | 1.97(1) |
| $^{11}$Li | 8.6 ms | $3/2^-$ | 2.482(14)(44) | 2.38(5) | |
| $^7$Be | 53 d | $3/2^-$ | 2.646(10)(16) | 2.507(17) | 2.47(1) |
| $^9$Be | *stable* | $3/2^-$ | 2.519(0)(12) | 2.385(13) | 2.37(1) |
| $^{10}$Be | 1.5 Myr | $0^+$ | 2.361(9)(17) | 2.224(18) | 2.19(1) |
| $^{11}$Be | 14 s | $1/2^+$ | 2.466(8)(15) | 2.341(16) | |
| $^{12}$Be | 24 ms | $0^+$ | 2.503(9)(15) | 2.386(16) | |

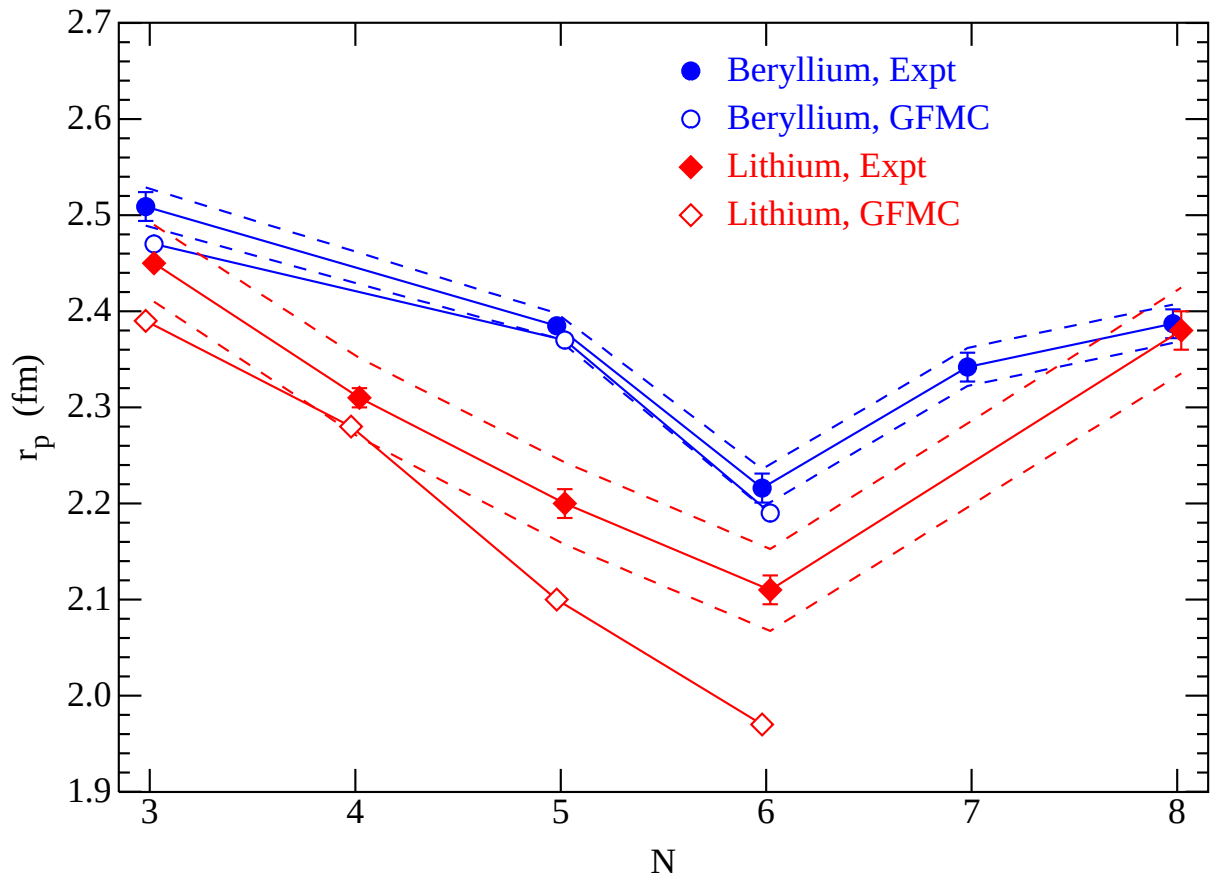

FIG. 15 Point proton radii of lithium and beryllium isotopes obtained from isotope shift measurements (Krieger *et al.*, 2012; Nörtershäuser *et al.*, 2011a). Error bars are based on the isotope shift uncertainty only. The additional systematic uncertainty caused by the reference charge radius uncertainty is indicated by the dashed lines. The GFMC values are for the AV18+IL7 Hamiltonian.

### B. Laser spectroscopic studies of $^{7,9,10,11,12}Be^+$

Laser spectroscopy has been performed on $^{7,9,10,11,12}Be^+$ produced by fragmentation in a uranium-carbide target induced by a 1.4 GeV proton beam at ISOLDE (Krieger *et al.*, 2012; Nörtershäuser *et al.*, 2009). The most neutron-rich among the bound beryllium isotopes, $^{14}$Be (half-life = 4.4 ms), remains a challenge due to its low yield of only a few ions per second at ISOL facilities. Among the isotopes studied, $^{12}$Be had the lowest yield: approximately 8000 $^{12}$Be nuclei were produced for a proton pulse that hits the target every 3–4 s. The beryllium ions were transported at an energy of 40–60 keV through a beamline where frequency-comb based collinear and anticollinear laser spectroscopy was performed on the $2s\,^2S_{1/2} - 2p\,^2P_{1/2,3/2}$ transitions. While collinear laser spectroscopy is applied for isotope shift measurements on-line already for about three decades, it was so far not possible to determine nuclear charge radii of isotopes lighter than neon. This is caused by the large uncertainty of the artificial isotope shift induced by the imprecisely known acceleration potential. The typical relative accuracy of the high-voltage measurement at the ISOLDE target and ion source platform is about $10^{-4}$. This corresponds to an uncertainty of about 40 MHz in the isotope shift between the stable isotope $^9$Be and $^{12}$Be, a factor of 40 larger than the required accuracy of about 1 MHz. The lightest short-lived isotope of which the charge radius was determined before by collinear spectroscopy was $^{17}$Ne in order to study a possible onset of a two-proton halo in this nucleus (Geithner *et al.*, 2008). To overcome this major technical difficulty, the absolute frequencies of both the collinear resonance ($f_c$) and the anticollinear resonance ($f_a$) were measured using a frequency comb. Therefore, two frequency-doubled dye lasers at 624 nm (collinear) and 628 nm (anticollinear) were locked to a reference transition in iodine and directly to the frequency comb, respectively. Resonance fluorescence was measured as a function of the acceleration voltage applied to the optical detection region. In order to suppress the background from scattered laser light during the measurement of the rarely produced $^{12}$Be, the cleanliness of the beam at mass $A = 12$ was exploited and a photon-ion delayed-coincidence technique applied to record only photons detected while a $^{12}Be^+$ ion was inside the optical detection region. From the fitted line centers the resonance frequencies in the laboratory system $f_c$ and $f_a$ were determined and the transition frequency in the rest frame of the ion was deduced as

$f_0 = \sqrt{f_c f_a}$.

Relative to the charge radius of $^9$Be — a value independently determined from electron scattering experiments — the nuclear charge radii of all the other Be isotopes were deduced (Table VIII). In addition, the magnetic moments of the unstable odd-mass nuclei were determined with improved accuracy. It should be mentioned that the beryllium isotopes $^{7,9,11}$Be were also investigated using laser-cooled ions captured in a linear Paul trap (Takamine *et al.*, 2009). While the magnetic dipole moment of $^7$Be was determined with high accuracy (Okada *et al.*, 2008), no final values have been reported for the charge radii so far.

The results obtained with the collinear technique are shown in Fig. 15 and are in good agreement with nuclear structure calculations using fermionic molecular dynamics and NCSM. The charge radius decreases monotonically along the $^7$Be–$^9$Be–$^{10}$Be isotope chain, probably due to the clusterization of $^7$Be into an $\alpha$+$^3$He cluster, whereas $^{9,10}$Be are considered to be $\alpha$+$\alpha$+n and $\alpha$+$\alpha$+n+n systems, respectively. The charge radius increases monotonically along $^{10}$Be–$^{11}$Be–$^{12}$Be. $^{11}$Be may be viewed as a two-body system consisting of a frozen $^{10}$Be core and a halo neutron, approximately 8 fm apart. The charge radius of $^{12}$Be provides important new information to understand its structure. In the fermionic molecular dynamics model, its large radius is related to a breakdown of the $N = 8$ shell closure with 70% admixture of the $(sd)^2$ state. The GFMC results using AV18+IL7 are in excellent agreement with the data.

## IX. OUTLOOK

The study of neutron-rich nuclei provides important insight into the nuclear forces that hold these loosely bound systems together. Until recently, there was no effective method to measure the nuclear charge radii for short-lived, light nuclei. Advances in both atomic structure theory and laser spectroscopy techniques have now changed this situation, leading to the precision measurements presented in this review. The results will serve as a proving ground to improve both the descriptions of the nuclear force and nuclear structure calculations.

There are more challenging cases for nuclear charge radius measurements that can be addressed by the laser spectroscopists: $^3$H ($t_{1/2} = 12$ yr), one of the two simplest three-body systems; $^{14}$Be ($t_{1/2} = 4.6$ ms), the most neutron-rich case among the known beryllium isotopes; $^8$B ($t_{1/2} = 770$ ms), the lightest isotope believed to exhibit an extended proton distribution.

For boron and elements of higher Z, the collinear/anticollinear approach developed for beryllium can in principle be used. However, without further advances on the theoretical side, spectroscopy has to be performed on multiply-charged ions, which makes the production of the ion beam and finding an appropriate transition difficult. Future advances on the high-precision variational techniques used in the atomic structure calculations could extend the results to include atoms with more than three electrons, so that a wider range of atomic species and ions can be studied by the isotope shift method. Moreover, calculations of even higher order QED terms could reduce the uncertainty of atomic structure theory to the point that the nuclear charge radius could be directly extracted from its effect on the atomic transition frequency, without the need for both the isotope shift method and for a reference isotope with a known radius. A solution to these problems would pave the way to even more profound advances at the interface between atomic and nuclear physics.

### Acknowledgments

We thank P. Maris for supplying results before publication. We thank W. Nazarewicz and T. P. O'Connor for supplying figures. We thank R. V. F. Janssens, D. C. Morton, and W. Nazarewicz for their helpful comments. Z.-T. Lu, P. Mueller and S. Pieper acknowledge the support by the Department of Energy, Office of Nuclear Physics, under contract DEAC02-06CH11357. W. Nörtershäuser acknowledges the support by the Helmholtz Association (contract VH-NG-148) and the BMBF (contract 06MZ7169). G. W. F. Drake and Z.-C. Yan acknowledge the support by NSERC, SHARCnet and ACEnet of Canada and by the CAS/SAFEA International Partnership Program for Creative Research Teams.

### References

Al-Khalili, J. S., J. A. Tostevin, and I. J. Thompson, 1996, Phys. Rev. C **54**, 1843.

Alkhazov, G. D., *et al.*, 1997, Phys. Rev. Lett. **78**, 2313.

Alkhazov, G. D., *et al.*, 2002, Nucl. Phys. A **712**, 269.

Antognini, A., *et al.*, 2011, Can. J. Phys. **89**, 47.

Antognini, A., *et al.*, 2013, Science **339**, 417.

Antonov, A. N., *et al.*, 2011, Nucl. Instrum. Meth. Phys. Res. A **637**, 60.

Barrett, B. R., P. Navrátil, and J. P. Vary, 2013, Prog. Part. Nucl. Phys. **69**, 131.

Baym, G., 1969, *Lectures on Quantum Mechanics* (Westview).

Beringer, J., J. F. Arguin, R. M. Barnett, K. Copic, O. Dahl, D. E. Groom, C. J. Lin, J. Lys, H. Murayama, C. G. Wohl, W. M. Yao, P. A. Zyla, *et al.* (Particle Data Group), 2012, Phys. Rev. D **86**, 010001, URL `http://link.aps.org/doi/10.1103/PhysRevD.86.010001`.

Bernauer, J., *et al.*, 2010, Phys. Rev. Lett. **105**, 242001.

Bethe, H. A., and E. E. Salpeter, 1957, *Quantum Mechanics of one- and two-electron atoms* (Springer-Verlag, New York).

Brida, I., S. C. Pieper, and R. B. Wiringa, 2011, Phys. Rev. C **84**, 024319, URL `http://link.aps.org/doi/10.1103/PhysRevC.84.024319`.

Brodeur, M., *et al.*, 2012, Phys. Rev. Lett. **108**, 052504.

Brown, R. C., S. Wu, J. V. Porto, C. J. Sansonetti, C. E. Simien, S. M. Brewer, J. N. Tan, and J. D. Gillaspy, 2013,

Phys. Rev. A **87**, 032504, URL http://link.aps.org/doi/10.1103/PhysRevA.87.032504.

Bushaw, B. A., W. Nörtershäuser, G. W. F. Drake, and H.-J. Kluge, 2007, Phys. Rev. A **75**, 052503, URL http://link.aps.org/doi/10.1103/PhysRevA.75.052503.

Cancio Pastor, P., L. Consolino, G. Giusfredi, P. De Natale, M. Inguscio, V. A. Yerokhin, and K. Pachucki, 2012, Phys. Rev. Lett. **108**, 143001, URL http://link.aps.org/doi/10.1103/PhysRevLett.108.143001.

Chen, C.-Y., Y. M. Li, K. Bailey, T. P. O'Connor, L. Young, and Z.-T. Lu, 1999, Science **286**, 1139.

Cingoz, A., *et al.*, 2012, Nature **482**, 68.

Cockrell, C., J. P. Vary, and P. Maris, 2012, Phys. Rev. C **86**, 034325, URL http://link.aps.org/doi/10.1103/PhysRevC.86.034325.

Collard, H., *et al.*, 1963, Phys. Rev. Lett. **11**, 132.

Dobrovolsky, A. G. A. M. B. A. E. P., A.V., S. Fritz, H. Geissel, C. Gross, A. Khanzadeev, and G. Korolev, 2006, Nucl. Phys. A **766**, 1.

Drake, G. W. F., 1993a, in *Long-range Casimir forces: theory and recent experiments on atomic systems*, edited by D. A. Micha and F. S. Levin (Plenum, New York), pp. 107–217.

Drake, G. W. F., 1993b, Adv. At. Mol. Opt. Phys. **31**, 1.

Drake, G. W. F., 1994, Adv. At. Mol. Opt. Phys. **32**, 93.

Drake, G. W. F., M. M. Cassar, and R. A. Nistor, 2002, Phys. Rev. A **65**, 054501.

Drake, G. W. F., and S. P. Goldman, 2000, Can. J. Phys. **77**, 835.

Drake, G. W. F., and W. C. Martin, 1999, Can. J. Phys. **77**, 835.

Drake, G. W. F., and Z.-C. Yan, 1992, Phys. Rev. A **46**, 2378.

Drake, G. W. F., and Z.-C. Yan, 1994, Chem. Phys. Lett. **229**, 486.

Epelbaum, E., 2006, Prog. Part. Nucl. Phys. **57**(2), 654 , ISSN 0146-6410, URL http://www.sciencedirect.com/science/article/pii/S0146641005001018.

Epelbaum, E., and U.-G. Meissner, 2012, Ann. Rev. Nucl. Part. Sci. **62**, 159 .

Esbensen, H., K. Hagino, P. Mueller, and H. Sagawa, 2007, Phys. Rev. C **76**, 024302, URL http://link.aps.org/doi/10.1103/PhysRevC.76.024302.

Ewald, G., W. Nörtershäuser, A. Dax, S. Götte, R. Kirchner, H.-J. Kluge, T. Kühl, R. Sanchez, A. Wojtaszek, B. A. Bushaw, G. W. F. Drake, Z.-C. Yan, *et al.*, 2004, Phys. Rev. Lett. **93**, 113002, URL http://link.aps.org/doi/10.1103/PhysRevLett.93.113002.

Firestone, R. B., and V. S. Shirley, 1996, *Table of Isotopes* (Wiley Interscience, New York).

Frederico, T., A. Delfino, L. Tomio, and M. Yamashita, 2012, Progress in Particle and Nuclear Physics **67**(4), 939 , ISSN 0146-6410, URL http://www.sciencedirect.com/science/article/pii/S0146641012000907.

Friar, J. L., J. Martorell, and D. W. L. Sprung, 1997, Phys. Rev. A **56**, 4579.

Fujita, J.-i., and H. Miyazawa, 1957, Progress of Theoretical Physics **17**(3), 360, URL http://ptp.ipap.jp/link?PTP/17/360/.

Geithner, W., T. Neff, G. Audi, K. Blaum, P. Delahaye, H. Feldmeier, S. George, C. Guénaut, F. Herfurth, A. Herlert, S. Kappertz, M. Keim, *et al.*, 2008, Phys. Rev. Lett. **101**, 252502, URL http://link.aps.org/doi/10.1103/PhysRevLett.101.252502.

Hagino, K., and H. Sagawa, 2005, Phys. Rev. C **72**, 044321, URL http://link.aps.org/doi/10.1103/PhysRevC.72.044321.

Hauser, P., *et al.*, 1992, Phys. Rev. A **46**, 2363.

Hylleraas, E. A., 1929, Zeitschrift fur Phys. **54**, 347.

Ilieva, S., F. Aksouh, G. D. Alkhazov, L. Chulkov, A. V. Dobrovolsky, P. Egelhof, H. Geissel, M. Gorska, A. Inglessi, R. Kanungo, A. V. Khanzadeev, O. A. Kiselev, *et al.*, 2012, Nucl. Phys. A **875**, 8.

Jensen, A. S., K. Riisager, D. V. Fedorov, and E. Garrido, 2004, Rev. Mod. Phys. **76**, 215, URL http://link.aps.org/doi/10.1103/RevModPhys.76.215.

Jentschura, U. D., 2011a, Ann. Phys. **326**, 500.

Jentschura, U. D., 2011b, Ann. Phys. **326**, 516.

Kandula, D. Z., C. Gohle, T. J. Pinkert, W. Ubachs, and K. S. E. Eikema, 2011, Phys. Rev. A **84**, 062512.

Karol, P. J., 1975, Phys. Rev. C **11**, 1203.

Kikuchi, Y., K. Katō, T. Myo, M. Takashina, and K. Ikeda, 2010, Phys. Rev. C **81**, 044308, URL http://link.aps.org/doi/10.1103/PhysRevC.81.044308.

Klahn, B., and W. A. Bingel, 1977, Theo. Chem. Acta **44**, 27.

Klahn, B., and W. A. Bingel, 1978, Int. J. Quantum Chem. **11**, 943.

Krieger, A., K. Blaum, M. L. Bissell, N. Frömmgen, C. Geppert, M. Hammen, K. Kreim, M. Kowalska, J. Krämer, T. Neff, R. Neugart, G. Neyens, *et al.*, 2012, Phys. Rev. Lett. **108**, 142501, URL http://link.aps.org/doi/10.1103/PhysRevLett.108.142501.

Landré-Pellemoine, F., J. Angelique, O. Bajeat, C. Baruė, R. Bennett, F. Clapier, M. Ducourtieux, G. Gaubert, S. Gibouin, Y. Huguet, P. Jardin, S. Kandri-Rody, *et al.*, 2002, Nuclear Physics A **701**(14), 491 , ISSN 0375-9474, 5th International Conference on Radioactive Nuclear Beams, URL http://www.sciencedirect.com/science/article/pii/S0375947401016335.

Machleidt, R., and D. Entem, 2011, Physics Reports **503**(1), 1 , ISSN 0370-1573.

Marin, F., *et al.*, 1995, Z. Phys. D **32**, 285.

Maris, P., 2013, private communication .

Maris, P., J. P. Vary, and A. M. Shirokov, 2009, Phys. Rev. C **79**, 014308, URL http://link.aps.org/doi/10.1103/PhysRevC.79.014308.

Metropolis, N., A. W. Rosenbluth, M. N. Rosenbluth, A. H. Teller, and E. Teller, 1953, J. Chem. Phys. **21**, 1087.

Mohr, P. J., B. N. Taylor, and D. B. Newell, 2012, Rev. Mod. Phys. **84**, 1527, URL http://link.aps.org/doi/10.1103/RevModPhys.84.1527.

Morton, D. C., Q. Wu, and G. W. F. Drake, 2006, Can. J. Phys. **84**, 83.

Mueller, P., L.-B. Wang, G. W. F. Drake, K. Bailey, Z.-T. Lu, and T. P. O'Connor, 2005, Phys. Rev. Lett. **94**, 133001.

Mueller, P., *et al.*, 2007, Phys. Rev. Lett. **99**, 252501.

Nakashima, H., and H. Nakatsuji, 2008, J. Chem. Phys. **128**, 154107.

Navrátil, P., J. P. Vary, and B. R. Barrett, 2000, Phys. Rev. Lett. **84**, 5728, URL http://link.aps.org/doi/10.1103/PhysRevLett.84.5728.

Neugart, R., D. L. Balabanski, K. Blaum, D. Borremans, P. Himpe, M. Kowalska, P. Lievens, S. Mallion, G. Neyens, N. Vermeulen, and D. T. Yordanov, 2008, Phys. Rev. Lett. **101**, 132502, URL http://link.aps.org/doi/10.1103/PhysRevLett.101.132502.

Nörtershäuser, W., T. Neff, R. Sánchez, and I. Sick, 2011a, Phys. Rev. C **84**, 024307, URL http://link.aps.org/doi/10.1103/PhysRevC.84.024307.

Nörtershäuser, W., R. Sánchez, G. Ewald, A. Dax, J. Behr, P. Bricault, B. A. Bushaw, J. Dilling, M. Dombsky, G. W. F. Drake, S. Götte, H.-J. Kluge, *et al.*, 2011b, Phys. Rev. A **83**, 012516, URL http://link.aps.org/doi/10.1103/PhysRevA.83.012516.

Nörtershäuser, W., D. Tiedemann, M. Žáková, Z. Andjelkovic, K. Blaum, M. L. Bissell, R. Cazan, G. W. F. Drake, C. Geppert, M. Kowalska, J. Krämer, A. Krieger, *et al.*, 2009, Phys. Rev. Lett. **102**, 062503, URL http://link.aps.org/doi/10.1103/PhysRevLett.102.062503.

Okada, K., M. Wada, T. Nakamura, A. Takamine, V. Lioubimov, P. Schury, Y. Ishida, T. Sonoda, M. Ogawa, Y. Yamazaki, Y. Kanai, T. M. Kojima, *et al.*, 2008, Phys. Rev. Lett. **101**, 212502, URL http://link.aps.org/doi/10.1103/PhysRevLett.101.212502.

Ong, A., J. C. Berengut, and V. V. Flambaum, 2010, Phys. Rev. C **82**, 014320.

Ottermann, C. R., G. Kobschall, K. Maurer, K. Rohrich, C. Schmitt, and V. H. Walther, 1985, Nucl. Phys. A **436**, 688.

Pachucki, K., and A. M. Moro, 2007, Phys. Rev. A **75**, 032521, URL http://link.aps.org/doi/10.1103/PhysRevA.75.032521.

Pachucki, K., and J. Sapirstein, 2003, J. Phys. B At. Mol. Opt. Phys. **36**, 803.

Pachucki, K., V. A. Yerokhin, and P. Cancio Pastor, 2012, Phys. Rev. A **85**, 042517, URL http://link.aps.org/doi/10.1103/PhysRevA.85.042517.

Papadimitriou, G., A. T. Kruppa, N. Michel, W. Nazarewicz, M. Płoszajczak, and J. Rotureau, 2011, Phys. Rev. C **84**, 051304, URL http://link.aps.org/doi/10.1103/PhysRevC.84.051304.

Pastore, S., S. C. Pieper, R. Schiavilla, and R. B. Wiringa, 2012, submitted to Phys. Rev. C URL http://arxiv.org/abs/1212.3375.

Pieper, S. C., 2005, Nuclear Physics A **751**(0), 516 , ISSN 0375-9474, ¡ce:title¿Proceedings of the 22nd International Nuclear Physics Conference (Part 1)¡/ce:title¿, URL http://www.sciencedirect.com/science/article/pii/S0375947405001247.

Pieper, S. C., 2008a, AIP Conference Proceedings **1011**(1), 143, URL http://link.aip.org/link/?APC/1011/143/1.

Pieper, S. C., 2008b, Proceedings of the International School of Physics "Enrico Fermi" **169**, 111.

Pieper, S. C., V. R. Pandharipande, R. B. Wiringa, and J. Carlson, 2001, Phys. Rev. C **64**, 014001, URL http://link.aps.org/doi/10.1103/PhysRevC.64.014001.

Pieper, S. C., and R. B. Wiringa, 2001, Annual Review of Nuclear and Particle Science **51**(1), 53.

Pieper, S. C., R. B. Wiringa, and J. Carlson, 2004, Phys. Rev. C **70**, 054325, URL http://link.aps.org/doi/10.1103/PhysRevC.70.054325.

Pohl, R., *et al.*, 2010, Nature **466**, 213.

Puchalski, M., D. Kedziera, and K. Pachucki, 2013, Phys. Rev. A **87**, 032503, URL http://link.aps.org/doi/10.1103/PhysRevA.87.032503.

Puchalski, M., A. M. Moro, and K. Pachucki, 2006, Phys. Rev. Lett. **97**, 133001.

Puchalski, M., and K. Pachucki, 2008, Phys. Rev. A **78**, 052511, URL http://link.aps.org/doi/10.1103/PhysRevA.78.052511.

van Rooij, R., J. S. Borbely, J. Simonet, M. D. Hoogerland, K. S. E. Eikema, R. A. Rozendaal, and W. Vassen, 2011, Science **333**, 196.

Ryjkov, V. L., *et al.*, 2008, Phys. Rev. Lett. **101**, 012501.

Sánchez, R., W. Nörtershäuser, G. Ewald, D. Albers, J. Behr, P. Bricault, B. A. Bushaw, A. Dax, J. Dilling, M. Dombsky, G. W. F. Drake, S. Götte, *et al.*, 2006, Phys. Rev. Lett. **96**, 033002, URL http://link.aps.org/doi/10.1103/PhysRevLett.96.033002.

Sánchez, R., M. Záková, Z. Andjelkovic, B. A. Bushaw, K. Dasgupta, G. Ewald, C. Geppert, H.-J. Kluge, J. Krämer, M. Nothhelfer, D. Tiedemann, D. F. A. Winters, *et al.*, 2009, New Journal of Physics **11**(7), 073016, URL http://stacks.iop.org/1367-2630/11/i=7/a=073016.

Sansonetti, C. J., C. E. Simien, J. D. Gillaspy, J. N. Tan, S. M. Brewer, R. C. Brown, S. Wu, and J. V. Porto, 2011, Phys. Rev. Lett. **107**, 023001, URL http://link.aps.org/doi/10.1103/PhysRevLett.107.023001.

Schwartz, C., 2006, Int. J. Mod. Phys. E **15**, 877.

Shiner, D., R. Dixson, and V. Vedantham, 1995, Phys. Rev. Lett. **74**, 3553.

Shirokov, A., J. Vary, A. Mazur, and T. Weber, 2007, Physics Letters B **644**(1), 33 , ISSN 0370-2693.

Sick, I., 2001, Progress in Particle and Nuclear Physics **47**(1), 245 .

Sick, I., 2008, Phys. Rev. C **77**, 041302(R).

Stone, A. P., 1961, Proc. Phys. Soc. (London) **77**, 786.

Stone, A. P., 1963, Proc. Phys. Soc. (London) **81**, 868.

Suda, T., and M. Wakasugi, 2005, Prog. Part. Nucl. Phys. **55**, 417.

Sulai, I. A., *et al.*, 2008, Phys. Rev. Lett. **101**, 173001.

Takamine, A., M. Wada, K. Okada, T. Nakamura, P. Schury, T. Sonoda, V. Lioubimov, H. Iimura, Y. Yamazaki, Y. Kanai, T. Kojima, A. Yoshida, *et al.*, 2009, Euro. Phys. J. A **42**, 369.

Tanihata, I., H. Savajols, and R. Kanungo, 2013, Prog. Part. Nucl. Phys. **68**(0), 215 , ISSN 0146-6410.

Tanihata, I., *et al.*, 1985a, Phys. Rev. Lett. **55**, 2676.

Tanihata, I., *et al.*, 1985b, Phys. Lett. B **160**, 380.

Tanihata, I., *et al.*, 1988, Phys. Lett. B **206**, 592.

Tanihata, I., *et al.*, 1992, Phys. Lett. B **289**, 261.

Vassen, W., *et al.*, 2012, Rev. Mod. Phys. **84**, 175.

Wang, L.-B., *et al.*, 2004, Phys. Rev. Lett. **93**, 142501.

Wang, M., *et al.*, 2012, Chinese Physics C **36**, 1603.

Wiringa, R. B., V. G. J. Stoks, and R. Schiavilla, 1995, Phys. Rev. C **51**, 38, URL http://link.aps.org/doi/10.1103/PhysRevC.51.38.

Yan, Z.-C., and G. W. F. Drake, 1994, Phys. Rev. A **50**, R1980.

Yan, Z.-C., and G. W. F. Drake, 1998, Phys. Rev. Lett. **81**, 774, URL http://link.aps.org/doi/10.1103/PhysRevLett.81.774.

Yan, Z.-C., and G. W. F. Drake, 2000, Phys. Rev. A **61**, 022504, URL http://link.aps.org/doi/10.1103/PhysRevA.61.022504.

Yan, Z.-C., W. Nörtershäuser, and G. W. F. Drake, 2008, Phys. Rev. Lett. **100**, 243002, URL http://link.aps.org/doi/10.1103/PhysRevLett.100.243002.

Zhan, X., *et al.*, 2011, Phys. Lett. B **705**, 59.

Zhukov, M. V., B. V. Danilin, D. V. Fedorov, J. M. Bang, I. J. Thompson, and J. S. Vaagen, 1993, Phys. Rep. **231**, 151.